\documentclass[twocolumn,reprint,superscriptaddress,preprintnumbers, amsmath,amssymb,aps,nofootinbib]{revtex4-1}

\usepackage{graphicx}
\usepackage{dcolumn}
\usepackage{bm}

\usepackage{hyperref}
\usepackage[utf8]{inputenc}
\usepackage[dvipsnames]{xcolor}
\usepackage[normalem]{ulem}
\usepackage{xspace}
\usepackage{fontawesome} 
\usepackage{multirow}

\usepackage[dvipsnames]{xcolor}
\usepackage[utf8]{inputenc}
\hypersetup{
    colorlinks=true,
    citecolor = green,
    linkcolor=red,
    filecolor=magenta,      
    urlcolor=blue,
}
\hypersetup{allcolors=[rgb]{0.0 0.0 0.9},linkcolor=[rgb]{0.75 0.05 0.05},urlcolor=[rgb]{0.6 0.0 0.6}, }

\usepackage{url}

\begin{document}

\preprint{BARI-TH/762-24}

\title{Status of tension between NOvA and T2K after Neutrino 2024\\
and  possible role of non-standard neutrino interactions}%

\author{		Sabya Sachi Chatterjee}
\email{sabya.chatterjee@kit.edu}
\affiliation{		Institut f\"{u}r Astroteilchenphysik, Karlsruher Institut f\"{u}r
Technologie (KIT), Hermann-von-Helmholtz-Platz 1, 76344
Eggenstein-Leopoldshafen, Germany}

\author{		Antonio Palazzo}
\email{palazzo@ba.infn.it}
\affiliation{ 	Dipartimento Interateneo di Fisica ``Michelangelo Merlin,'' Via Amendola 173, 70126 Bari, Italy}
\affiliation{ 	Istituto Nazionale di Fisica Nucleare, Sezione di Bari, Via Orabona 4, 70126 Bari, Italy}


\begin{abstract}

In a previous work we have shown that the data presented by the two long-baseline accelerator experiments NOvA and T2K at the Neutrino 2020 conference displayed a tension, and that it could be alleviated by non-standard neutrino interactions (NSI) of the flavor changing type involving the 
$e-\mu$ or the $e-\tau$ sectors with couplings $|\varepsilon_{e\mu}| \sim |\varepsilon_{e\tau}|\sim 0.1$. As a consequence a hint in favor of NSI emerged. In the present paper we reassess the issue in light of the new data released by the two experiments at the Neutrino 2024 conference. We find that the tension in the determination of the standard CP-phase $\delta_{\mathrm {CP}}$ extracted by the two experiments in the normal neutrino mass ordering persists and has a statistical significance of $\sim2\sigma$.
Concerning the NSI, we find that including their effects in the fit, the two values of  $\delta_{\mathrm {CP}}$ preferred by NOvA and T2K 
return in very good agreement. The current statistical significance of the hint of non zero NSI is $\sim1.8\sigma$.  
Further experimental data are needed in order to settle the issue.

\end{abstract}
\maketitle

{\bf {\em Introduction.}}  The two long-baseline (LBL) accelerator experiments NOvA and T2K have recently released new
data at the Neutrino 2024 Conference~\cite{NOVA_talk_nu2024,T2K_talk_nu2024}. Interestingly, the results in the $\nu_\mu \to \nu_e$ appearance channel of the two experiments continue to be in tension confirming the trend observed in previous data released 
at the Neutrino 2020 Conference~\cite{NOVA_talk_nu2020,T2K_talk_nu2020} (subsequently published in~\cite{NOvA:2021nfi,T2K:2023smv}),
and point towards values of the CP-phase $\delta_{\mathrm {CP}}$ which are in disagreement when the data are interpreted in the standard 3-flavor framework for normal ordered neutrino mass eigenstates.%
\footnote{In the 3-flavor framework one introduces three mass eigenstates $\nu_i$ with masses
$m_i\, (i = 1,2,3)$, three mixing angles $\theta_{12},\theta_{13}, \theta_{13}$, and one CP-phase $\delta_{\mathrm {CP}}$. The neutrino mass ordering (NMO) is said to be normal (inverted) if $m_3>m_{1,2}$  ($m_3<m_{1,2}$). We will abbreviate normal (inverted) ordering as NO (IO).}

 {\bf {\em Quantification of Tension.}} The mismatch between the preferred values of the $\delta_{\mathrm {CP}}$ is clear,
with T2K preferring a value of $\delta_{\mathrm {CP}}\simeq 1.5\pi$, and NOvA indicating
$\delta_{\mathrm {CP}}\simeq 0.9\pi$ (see~\cite{NOVA_talk_nu2024,T2K_talk_nu2024}).
Let's try to quantify the tension. Following~\cite{Maltoni:2003cu}, we introduce the function
 $\bar\chi^2(\delta_{\mathrm {CP}}) = \chi^2_{\rm{T2K+NOvA}}(\delta_{\mathrm {CP}}) - (\chi^2_{\rm{T2K} ,min } +\chi^2_{\rm{NOvA},min})$.
 Figure~\ref{fig:chi2-plot_NO} 
shows the function $\bar \chi^2(\delta_{\mathrm {CP}})$
together with the two functions  $\Delta\chi^2_{r}(\delta_{\mathrm {CP}}) = \chi^2_{r}(\delta_{\mathrm {CP}}) - \chi^2_{r,min }$,
where $r$ is an index designating the experiment in question (T2K or NOvA).
The level of compatibility between the two experiments can be quantified by  means of the minimum value  $\bar\chi^2_{min} \simeq 6.3$,
which for 2 d.o.f.%
\footnote{Note that the only two relevant d.o.f. are $\delta_{\mathrm {CP}}$ and $\theta_{23}$.
In fact, the  parameters $\theta_{12}$, $\Delta m^2_{21}$ can be considered fixed by solar neutrinos and KamLAND,
$\theta_{13}$ is fixed by Daya Bay, and $\Delta m^2_{31}$  is fixed with high precision by 
the disappearance channel measurements of T2K and NOvA themselves. Concerning this
last parameter it is useful to notice that the estimate of  $\Delta m^2_{31}$  provided by the
disappearance channel is completely insensitive to the couplings  $\varepsilon_{e\mu}$ and  $\varepsilon_{e\tau}$ within their range of interest, 
as we have explicitly checked numerically (see the right panels of Fig.~S1 in the Supplemental Material).}
corresponds to a goodness-of-fit (GoF) of $4.3 \times 10^{-2}$ (equivalent to $2.0\sigma$).

 An alternative method to quantify the tension is to compare the estimates of $\delta_{\mathrm {CP}}$
 given by the two experiments. For T2K and NOvA we assume that the errors are Gaussian. As suggested
 by Fig.~\ref{fig:chi2-plot_NO} this assumption is reasonable for T2K, while for NOvA it is valid only for
 the upper error, which is the relevant one for estimating the tension.
 We have for T2K $\delta_{\mathrm{CP}}/\pi = 1.47 \pm  0.24$, while for NOvA  $\delta_{\mathrm {CP}}/\pi = 0.87^{+0.20}_{-0.19}$.
 Considering the two errors summed in quadrature we have that the two estimates differ by  $\Delta \delta_{\mathrm {CP}}/\pi = 0.60 \pm 0.31$,
 hence corroborating the $2.0\sigma$  level of tension found with the first criterion.%
\footnote{It is interesting to note that, as pointed out in~\cite{Blennow:2014sja}, due to the cyclic nature of the $\delta_{\mathrm {CP}}$
parameter (which implies the violation of the hypotheses underlying the Wilk's  theorem that become more apparent when the experiments have poor sensitivity to $\delta_{\mathrm {CP}}$), the real statistical level of the tension 
could be higher than that obtained using the $\chi^2$ estimator or gaussian errors.
An educated guess based on the numerical simulations performed in~\cite{Blennow:2014sja} (see Fig.~1), is that
a more faithful estimate of the statistical significance of the T2K-NOvA tension should lie somewhere in the interval $[2.0\sigma, 2.5\sigma]$.}
\begin{figure}[t!]
\vspace*{-0.4cm}
\hspace*{-0.2cm}
\includegraphics[height=7.2cm,width=8.7cm]{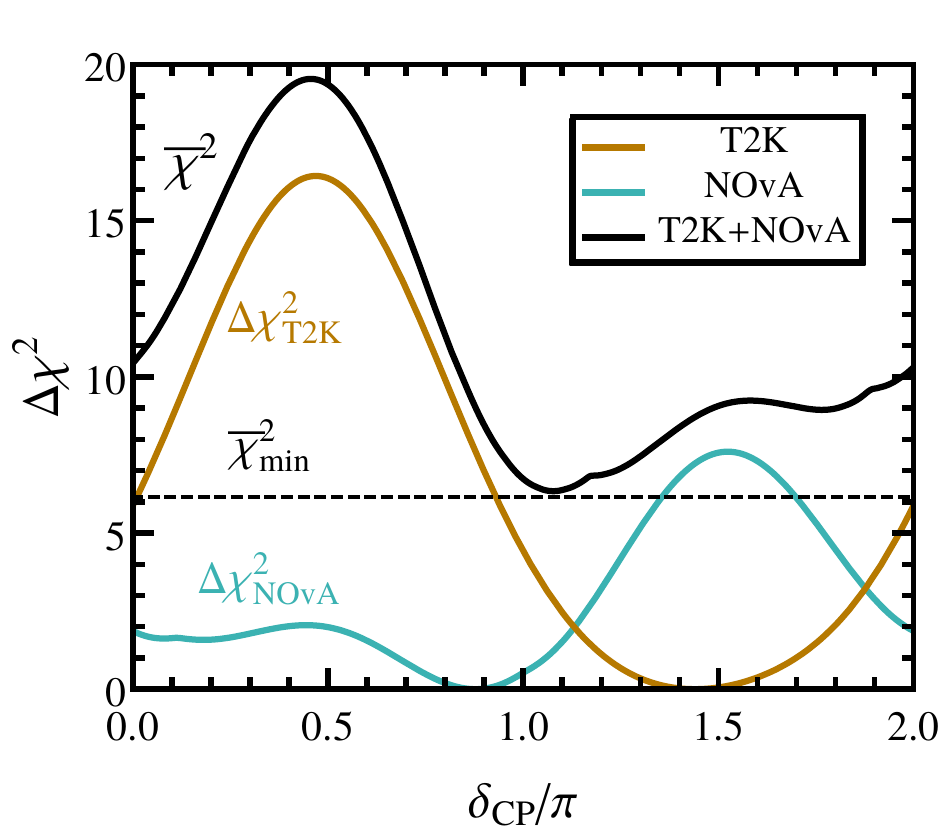}
\caption{Plot of the functions $\Delta\chi^2_{\rm{T2K}}$, $\Delta\chi^2_{\rm{NOvA}}$ and  $\bar\chi^2$ 
 as a function of $\delta_{\mathrm {CP}}$ for normal ordering.}
\label{fig:chi2-plot_NO}
\end{figure} 

{\bf {\em Motivation for NSI.}} While the observed discrepancy may be induced by a statistical fluctuation or an unknown systematic error, it
can represent a hint of new physics beyond the Standard Model (SM). As we already underlined in our previous paper~\cite{Chatterjee:2020kkm}, NOvA and T2K represent the ideal place where to seek non-standard matter effects in neutrino propagation due to their different and complementary setups. In particular, the two experiments work at two different peak energies
 (2 GeV for NOvA and 0.6 GeV for T2K) because of  the different baselines (810 km for NOvA and 295 km for T2K). As a consequence, in NOvA  matter effects are approximately three times larger with respect to T2K. 
  
 In~\cite{Chatterjee:2020kkm} (see also~\cite{Denton:2020uda}) we pointed out that the discrepancy could be solved by hypothesising the existence of NSI. More specifically,  we found a $\sim 2\sigma$ level preference for non-zero complex neutral-current (NC) NSI of the flavor changing type involving the $e-\mu$ or the $e-\tau$ sectors with couplings $|\varepsilon_{e\mu}| \sim |\varepsilon_{e\tau}|\sim 0.1$.%
\footnote{More precisely  the statistical significance of the indication we found in our previous paper~\cite{Chatterjee:2020kkm} was 
at the 2.1$\sigma$ level for $e-\mu$  NSI and 1.9$\sigma$ level fro $e-\tau$ NSI.}
 In the present paper we reassess the tension issue, showing that the indication persists in the new data with a statistical significance of $1.8\sigma$.%
 
{\bf {\em Theoretical framework.}} NSI may serve as the low-energy manifestation of high-energy physics, potentially arising from new, heavy states (for a review, see~\cite{Farzan:2017xzy,Biggio:2009nt,Ohlsson:2012kf,Miranda:2015dra,Dev:2019anc}). Alternatively, and perhaps more intriguingly, they could be linked to light mediators~\cite{Farzan:2015doa,Farzan:2015hkd}. As first pointed out in~\cite{Wolfenstein:1977ue}, NSI can drastically modify the dynamics of neutrino flavor conversion while passing through matter~\cite{Wolfenstein:1977ue,Mikheev:1986gs,Mikheev:1986wj}.
The influence of NSI on current and upcoming long-baseline (LBL) neutrino experiments has been extensively studied (see for example~\cite{Friedland:2012tq,Coelho:2012bp,Girardi:2014kca,Rahman:2015vqa,Coloma:2015kiu,deGouvea:2015ndi,Agarwalla:2016fkh,Liao:2016hsa,Forero:2016cmb,Huitu:2016bmb,Bakhti:2016prn,Masud:2016bvp,Soumya:2016enw,Masud:2016gcl,deGouvea:2016pom,Fukasawa:2016lew,Liao:2016orc,Liao:2016bgf,Blennow:2016etl,Deepthi:2017gxg,Flores:2018kwk,Hyde:2018tqt,Masud:2018pig,Esteban:2019lfo,Capozzi:2019iqn,Chatterjee:2021wac}). In particular, the tension observed between the T2K and NOvA experiments has prompted numerous investigations into NSI~\cite{Brahma:2023wlf,Cherchiglia:2023ojf,NOvA:2024lti} and other new physics scenarios, such as sterile neutrinos~\cite{Chatterjee:2020yak,deGouvea:2022kma}, non-unitary mixing~\cite{Miranda:2019ynh,Forero:2021azc}, violations of Lorentz Invariance~\cite{Rahaman:2021leu}, vector leptoquarks~\cite{Majhi:2022wyp}, ultra-light dark matter~\cite{Lin:2023xyk}, Leggett-Garg inequality violations~\cite{Konwar:2024nwc}, and dark photons~\cite{Alonso-Alvarez:2024wnh}. For a comprehensive review of these various scenarios, see~\cite{Rahaman:2022rfp}.

The NSI of neutral current (NC) type can be expressed in terms of a dimension-six operator~\cite{Wolfenstein:1977ue}:
\begin{equation}
\mathcal{L}_{\mathrm{NC-NSI}} \;=\;
-2\sqrt{2}G_F 
\varepsilon_{\alpha\beta}^{fC}
\bigl(\overline{\nu}_{\alpha}\gamma^\mu P_L \nu_\beta\bigr)
\bigl(\overline{f}\gamma_\mu P_C f\bigr)
\;,
\label{H_NC-NSI}
\end{equation}
where $\alpha, \beta = e, \mu, \tau$ represent the neutrino flavors, and $f = e, u, d$ refer to the matter fermions. The projector operator $P$ has a subscript $C=L, R$, which denotes the chirality of the fermion current, while $\varepsilon_{\alpha\beta}^{fC}$ are the NSI coupling amplitudes. The hermiticity of the Hamiltonian imposes the condition:
\begin{equation}
\varepsilon_{\beta\alpha}^{fC} \;=\; (\varepsilon_{\alpha\beta}^{fC})^*
\;.
\end{equation}

For neutrinos propagating through the Earth, one can define the effective strength of the NSI couplings as
\begin{equation}
\varepsilon_{\alpha\beta}
\;\equiv\; 
\sum_{f=e,u,d}
\varepsilon_{\alpha\beta}^{f}
\dfrac{N_f}{N_e}
\;\equiv\;
\sum_{f=e,u,d}
\left(
\varepsilon_{\alpha\beta}^{fL}+
\varepsilon_{\alpha\beta}^{fR}
\right)\dfrac{N_f}{N_e}
\;,
\label{epsilondef}
\end{equation}
$N_f$ being the number density of $f$ fermion.
Since the Earth matter can be considered as neutral and isoscalar, with  $N_n \simeq N_p = N_e$, 
we have $N_u \simeq N_d \simeq 3N_e$.
Hence,
\begin{equation}
\varepsilon_{\alpha\beta}\, \simeq\,
\varepsilon_{\alpha\beta}^{e}
+3\,\varepsilon_{\alpha\beta}^{u}
+3\,\varepsilon_{\alpha\beta}^{d}
\;.
\label{epsilon_eff}
\end{equation}
The NSI modify the effective Hamiltonian of neutrino  
in matter, which in the flavor basis can be expressed as
\begin{equation}
H \;=\; 
U
\begin{bmatrix} 
0 & 0 & 0 \\ 
0 & k_{21}  & 0 \\ 
0 & 0 & k_{31} 
\end{bmatrix}
U^\dagger
+
V_{\mathrm{CC}}
\begin{bmatrix}
1 + \varepsilon_{ee}  & \varepsilon_{e\mu}      & \varepsilon_{e\tau}   \\
\varepsilon_{e\mu}^*  & \varepsilon_{\mu\mu}    & \varepsilon_{\mu\tau} \\
\varepsilon_{e\tau}^* & \varepsilon_{\mu\tau}^* & \varepsilon_{\tau\tau}
\end{bmatrix}\,,
\end{equation}
where $U$ is the Pontecorvo-Maki-Nakagawa-Sakata (PMNS) matrix, which
consists of three mixing angles ($\theta_{12}, \theta_{13}, \theta_{23}$) and the CP-phase $\delta_{\mathrm {CP}}$.
We have denoted with $k_{21} \equiv \Delta m^2_{21}/2E$ and $k_{31} \equiv \Delta m^2_{31}/2E$ the solar and atmospheric wavenumbers respectively, with $\Delta m^2_{ij} \equiv m^2_i-m^2_j$, while
$V_{\mathrm{CC}}$ represnts the charged-current matter potential 
\begin{equation}
V_{\mathrm{CC}} 
\;=\; \sqrt{2}G_F N_e 
\;\simeq\; 7.6\, Y_e \times 10^{-14}
\bigg[\dfrac{\rho}{\mathrm{g/cm^3}}\bigg]\,\mathrm{eV}\,,
\label{matter-V}
\end{equation}
where $Y_e = N_e/(N_p+N_n) \simeq 0.5$ is the relative electron number density in the Earth crust.
To facilitate the analysis of matter effects, it turns out to be useful to introduce the dimensionless parameter $v = V_{\mathrm{CC}}/k_{31}$, which 
gauge the sensitivity to matter effects. The magnitude of this parameter is given by:
\begin{equation}
|v| 
\;=\; \bigg|\frac{V_{\mathrm{CC}}}{k_{31}}\bigg| 
\;\simeq\; 8.8 \times 10^{-2} \bigg[\frac{E}{\mathrm{GeV}}\bigg]\;,
\label{matter-v}
\end{equation}
and it prominently features in the analytical form of the $\nu_\mu \to \nu_e$ conversion probability.
Notably, in the two experiments, the first oscillation peak occurs at different energies with $E \simeq 0.6\, {\mathrm{GeV}}$ for T2K and $E \simeq 2\, {\mathrm{GeV}}$ for NOvA. Consequently, the matter effects in NOvA are approximately three times stronger ($v \simeq0.18$) than in T2K ($v \simeq 0.05$). As previously discussed in~\cite{Chatterjee:2020kkm}, this heightened sensitivity makes NOvA particularly responsive to NSI, while T2K remains largely unaffected. This disparity may explain the apparent tension between the two experiments when interpreted within the standard 3-flavor framework, ignoring NSI contributions.

In line with our earlier study~\cite{Chatterjee:2020kkm}, we focus on flavor non-diagonal NSI, where $\varepsilon_{\alpha\beta}$ for $\alpha\ne\beta$ plays a central role. Importantly, only these flavor changing NSI introduce dependence on a new CP-violating phase, which could be key to resolving the observed anomaly between NOvA and T2K. Specifically, we consider the parameters $\varepsilon_{e\mu}$ and $\varepsilon_{e\tau}$, which, as we will show below, induce an additional CP-phase dependence in the $\nu_{\mu} \to \nu_e$ transition probability%
\footnote{The $\nu_\mu \to \nu_\mu$ disappearance channel is sensitive to the ${\mu-\tau}$ NSI but 
this can be safely ignored because of the very strong upper bound
put with the atmospheric neutrinos by ANTARES~\cite{ANTARES:2021crm}, IceCube~\cite{IceCube:2022ubv} and 
KM3NeT/ORCA6~\cite{Lazo:2023bvy}, which all indicate $|\varepsilon_{\mu\tau}| \lesssim 5 \times 10^{-3}$ 
(see also~\cite{Aartsen:2017xtt} for a bound from SuperKamiokande). Interestingly, all the three experiments
find that $\varepsilon_{\mu\tau}=0$ is disfavored slightly below the 90\% C.L.}.
 Let us now consider the transition probability relevant for the T2K and NOvA experiments. When accounting for NSI, the probability can be expressed as the sum of three terms~\cite{Kikuchi:2008vq}: 
\begin{eqnarray}
\label{eq:Pme_4nu_3_terms}
P_{\mu e}  \simeq  P_{\rm{0}} + P_{\rm {1}}+   P_{\rm {2}}\,,
\end{eqnarray}
which, making use of a notation first introduced in~\cite{Liao:2016hsa}, take the expressions
\begin{eqnarray}
\label{eq:P0}
 & P_{\rm {0}} &\,\, \simeq\,  4 s_{13}^2 s^2_{23}  f^2\,,\\
\label{eq:P1}
 & P_{\rm {1}} &\,\,  \simeq\,   8 s_{13} s_{12} c_{12} s_{23} c_{23} \alpha f g \cos({\Delta + \delta_{\mathrm {CP}}})\,,\\
 \label{eq:P2}
 & P_{\rm {2}} &\,\,  \simeq\,  8 s_{13} s_{23} v |\varepsilon|   
 [a f^2 \cos(\delta_{\mathrm {CP}} + \phi) + b f g\cos(\Delta + \delta_{\mathrm {CP}} + \phi)]\,,\nonumber\\
\end{eqnarray}
$\Delta \equiv  \Delta m^2_{31}L/4E$ being the atmospheric oscillating factor,
$L$ is the baseline and $E$ the neutrino energy, and $\alpha \equiv \Delta m^2_{21}/ \Delta m^2_{31}$.
For compactness, we denote ($s_{ij} \equiv \sin \theta_{ij} $, $c_{ij} \equiv \cos \theta_{ij}$), 
and following~\cite{Barger:2001yr}, we introduce 
\begin{eqnarray}
\label{eq:S}
f \equiv \frac{\sin [(1-v) \Delta]}{1-v}\,, \qquad  g \equiv \frac{\sin v\Delta}{v}\,.
\end{eqnarray}
In Eq.~(\ref{eq:P2}) we have considered for the NSI coupling the complex form
\begin{eqnarray}
\varepsilon_{\alpha \beta} = |\varepsilon_{\alpha \beta}|  e^{i\phi_{\alpha \beta}}\,.
\end{eqnarray}
Notably, the form of $P_2$ is different for $\varepsilon_{e\mu}$ and  $\varepsilon_{e\tau}$ and,
in Eq. (\ref{eq:P2}), one has to perform the substitutions
\begin{eqnarray}
 \label{eq:P2_NSI_1}
 a = s^2_{23}, \quad b = c^2_{23} \quad &{\mathrm {if}}& \quad \varepsilon = |\varepsilon_{e\mu}|e^{i{\phi_{e\mu}}}\,,\\
  \label{eq:P2_NSI_2}
 a =  s_{23}c_{23}, \quad b = -s_{23} c_{23} \quad &{\mathrm {if}}& \quad \varepsilon = |\varepsilon_{e\tau}|e^{i{\phi_{e\tau}}}\,.
\end{eqnarray}
In the expressions given in Eqs.~(\ref{eq:P0})-(\ref{eq:P2}), 
the sign of $\Delta$, $\alpha$ and $v$ is positive (negative) for NO (IO).  
The expressions of the probability given above are valid for
neutrinos and the corresponding formulae for antineutrinos can be derived
by inverting in Eqs.~(\ref{eq:P0})-(\ref{eq:P2}) the sign of all the CP-phases and of $v$. 
Finally, we notice that the third term $P_{\rm {2}}$  depends
on the (complex) NSI coupling and it is non-zero only in the presence of matter (i.e. if $v \ne 0$). 
Physically, it originates from the interference of the matter potential
 $\varepsilon_{e\mu}V_{CC}$  (or $\varepsilon_{e\tau}V_{CC}$)
 with the atmospheric wavenumber $k_{31}$ (see~\cite{Friedland:2012tq}).%

{\bf {\em  Data used in the analysis.}}  We have made use of the datasets for the NOvA and T2K experiments from the most recent data releases, as presented in~\cite{NOVA_talk_nu2024} and~\cite{T2K_talk_nu2024}. In our analysis, we have fully accounted for both the disappearance and appearance channels for each experiment. For the numerical simulations, we use the GLoBES software package~\cite{Huber:2004ka,Huber:2007ji} along with an additional public tool~\cite{Kopp_NSI} designed to implement non-standard interactions. To perform the analysis, we marginalized over $\theta_{13}$ using a 2.8\% 1 sigma prior, with a central value of $\sin^2\theta_{13}= 0.0218$, as determined by the Daya Bay experiment~\cite{DayaBay:2022orm}. The solar parameters $\Delta m^2_{21}$ and $\theta_{12}$ are taken at their best fit values found in the global fit~\cite{Capozzi:2021fjo}.

{\bf {\em Numerical Results.}}  In Fig.~\ref{fig:regions_emu_NO} we present the numerical results 
obtained in the case of NO by combining NOvA and T2K. The three panels represent the projections
in the planes spanned by each pair of the three parameters $|\varepsilon_{e\mu}|$,
$\phi_{e\mu}$ and $\delta_{\mathrm {CP}}$. In all plots, the undisplayed parameters 
($\theta_{23}$, $\theta_{13}$ and $\Delta m^2_{31}$) are marginalized. 
The regions displayed are those allowed at the 68\% and 90\% confidence level for 2 d.o.f.
The first projection shows that both standard and non-standard
CP-phases have best fit values around $1.5\pi$, with the standard CP-phase  $\delta_{\mathrm {CP}}$
being more constrained. The second and third panels represent the projections spanned by the NSI coupling
and one of the two CP-phases. From these two plots we observe that there is a preference for
a NSI coupling different from zero with best fit  $|\varepsilon_{e\mu}| =  0.125$ and statistical significance
$\Delta \chi^2 = 3.1$ (corresponding to 1.76$\sigma$ for 1 d.o.f.).
Figure~\ref{fig:regions_etau_NO} is analogous to Fig.~\ref{fig:regions_emu_NO}. In this case, 
however, the coupling considered is $\varepsilon_{e\tau}$ (with the associated CP-phase $\phi_{e\tau}$).
The behaviour of the CP-phases is similar to the previous case. The values preferred
for the NSI coupling are larger with best fit $|\varepsilon_{e\tau}| =  0.22$. In this case,
the statistical significance of the preference of non-zero NSI coupling is $\Delta \chi^2 = 3.2$ (corresponding to 1.79$\sigma$ for 1 d.o.f.).

In Table~\ref{table:chi2} are reported the best fit values of the NSI couplings, those of the two 
CP-phases and what we obtain for  $\Delta\chi^2=\chi^2_{\rm SM}-\chi^2_{\rm SM + NSI}$ for each of the two 
possible choices of the neutrino mass orderings.%
\footnote{Note that the best fit values we obtain for the complex NSI couplings $\varepsilon_{e\mu}$ and $\varepsilon_{e\tau}$ 
are almost pure imaginary (being $\phi_{e\mu}$ and $\phi_{e\tau}$ close to $\sim 1.5\pi$). For this reason
 they cannot be confronted with the results of the NSI global analysis~\cite{Coloma:2023ixt}, 
where only real NSI couplings are considered. We may only observe qualitatively that the size
 $|\varepsilon_{e\mu}| =  0.125$ and $|\varepsilon_{e\tau}| =  0.22$ are compatible with the 90\% C.L. reported in~\cite{Coloma:2023ixt}.}
It is interesting to estimate at what statistical significance the SM  hypothesis is rejected. 
For NO, in the case of $e-\mu$ and $e-\tau$ NSI we obtain $\Delta\chi^2= 3.1$ and $\Delta\chi^2= 3.2$ respectively, which 
correspond (considering 2 d.o.f.) to an exclusion close to the $1.3 \sigma$ level.  
\begin{figure}[t!]
\vspace*{-0.8cm}
\hspace*{-0.2cm}
\includegraphics[height=9.5cm,width=9.5cm]{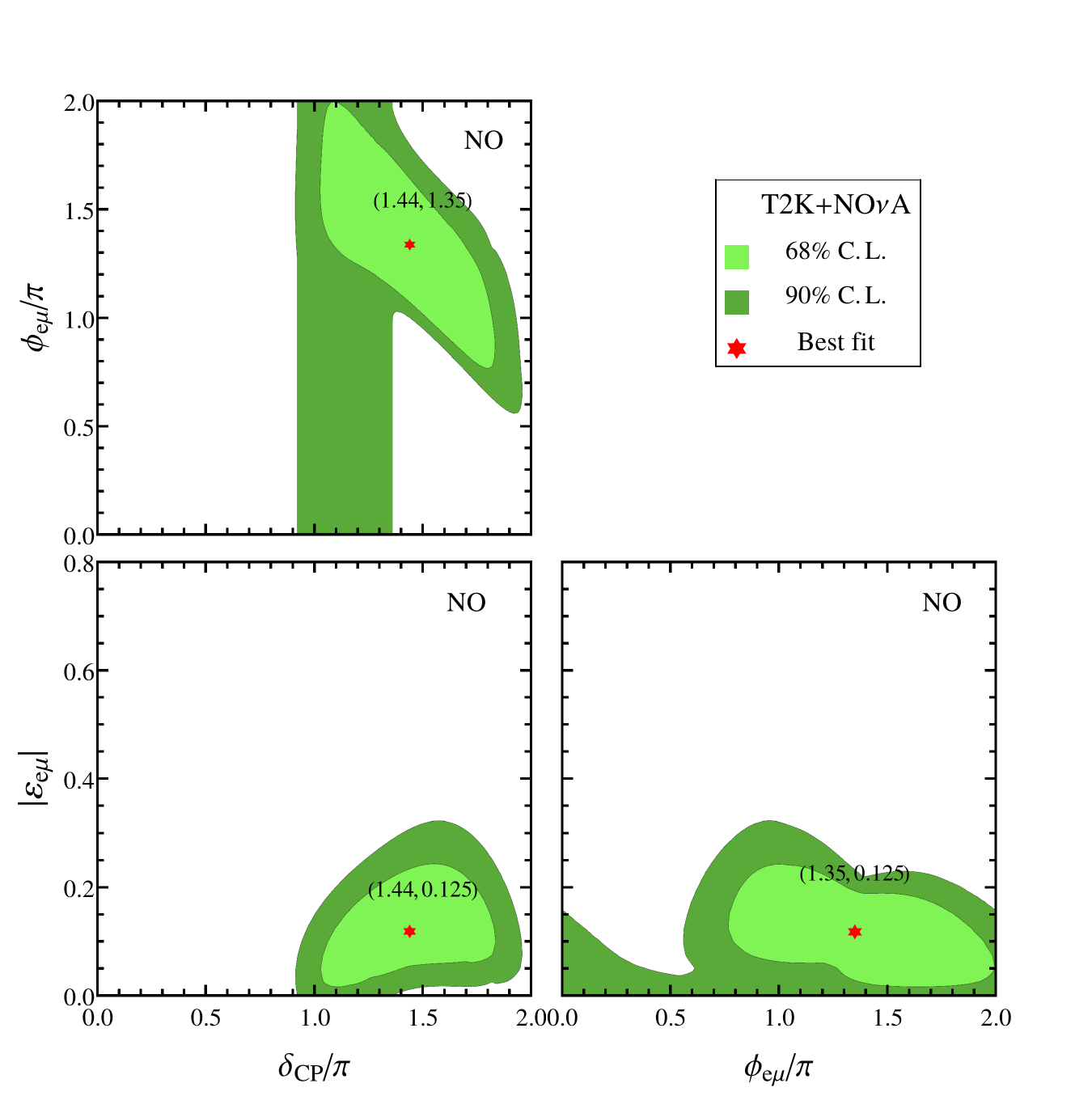}
\caption{Allowed regions determined by the combination of T2K and NOvA in NO for NSI of the $e-\mu$ type.}
\label{fig:regions_emu_NO}
\end{figure} 

With the  purpose of clarifying how the hint of a non-zero NSI coupling emerges, it is helpful 
to consider separately the two experiments NOvA and T2K. In 
Fig.~\ref{fig_tension_NO} we show the allowed regions for the NO case, 
in the plane of the two parameters $\delta_{\mathrm {CP}}$ and $\theta_{23}$.
The left panel depicts the SM scenario, while the central and right panels represent 
the SM+NSI cases with NSI of the $e-\mu$ and $e-\tau$ type respectively.
We underline that the SM regions in the left panel are in excellent agreement with those shown in the 
official plots of the collaborations~\cite{NOVA_talk_nu2024,T2K_talk_nu2024}, hence testifying the high level of accuracy reached 
by our analysis. In the central and right panels the NSI parameters are fixed at the best fit 
obtained from the combination of NOvA and T2K. These correspond to 
$|\varepsilon_{e\mu}| = 0.125, \phi_{e\mu} = 1.35\pi$ (central panel)
and $|\varepsilon_{e\tau}| = 0.22, \phi_{e\tau} = 1.70\pi$ (right panel). 
The two contours refer to the 68$\%$ and 90$\%$ C.L. for 2 d.o.f. 
In the SM case (left plot) a discrepancy between the values of $\delta_{\mathrm {CP}}$
 identified by the two experiments appears in a clear way, as already discussed
 above when commenting Figure~\ref{fig:chi2-plot_NO}. 
The diminishment of the discrepancy among the two experiments attained when NSI  are present 
is evident both in the central and right panels where there is a high level of overlap 
of the allowed regions for values of $\delta_{\mathrm {CP}}$ 
close to $1.5\pi$. The different behavior between the central and the right panel is due
to the different expression of transition probability for NSI of the $e-\mu$ or  $e-\tau$ type.
We can observe that the T2K regions are almost unaltered in the
presence of NSI (due to low sensitivity of its setup to matter effects). Differently,
the NOvA regions drastically change in the presence of NSI, due to the higher sensitivity
to matter effects. These findings are in line with our analytical discussion and underline
the high level of synergy and complementarity of the two setups.

\begin{figure}[t!]
\vspace*{-0.8cm}
\hspace*{-0.2cm}
\includegraphics[height=9.5cm,width=9.5cm]{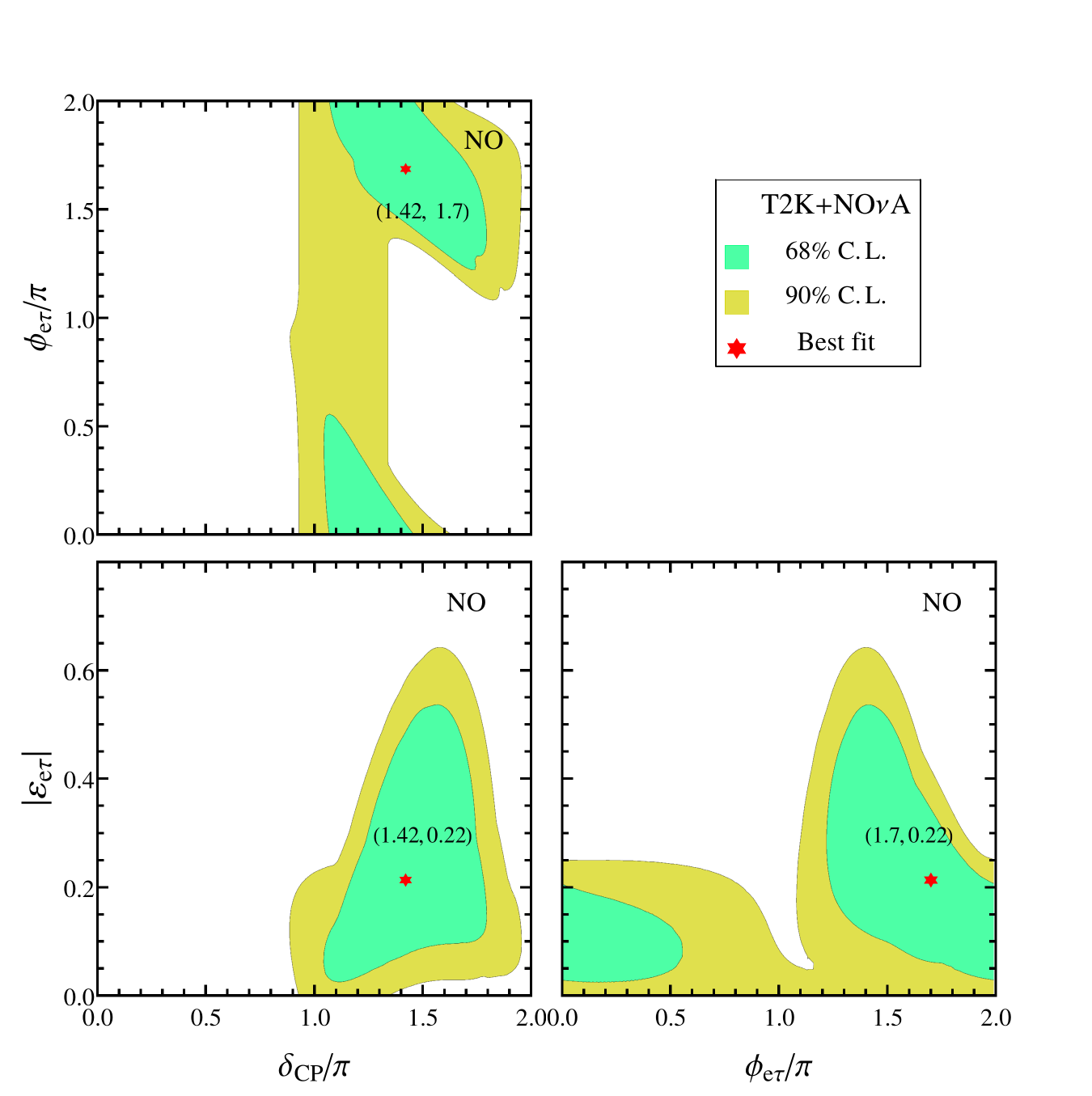}
\caption{Allowed regions determined by the combination of T2K and NOvA in NO for NSI of the $e-\tau$ type.}
\label{fig:regions_etau_NO}
\end{figure} 

\begin{table}[b!]
\centering
\caption{Best fit values and $\Delta\chi^2=\chi^2_{\rm SM}-\chi^2_{\rm SM+NSI}$ for the two choices of the NMO.}
\vspace{0.5cm}
\begin{tabular}{c|c|c|c|c|c}
NMO&NSI&$|\varepsilon_{\alpha\beta}|$&$\phi_{\alpha\beta}/\pi$&$\delta_{\mathrm {CP}}/\pi$&$\Delta\chi^2$\\\hline
\multirow{2}{*}{NO}&$\varepsilon_{e\mu}$&0.13&1.35&1.44&3.1\\
&$\varepsilon_{e\tau}$&0.22&1.70&1.42&3.2\\\hline
\multirow{2}{*}{IO}&$\varepsilon_{e\mu}$&0.05&1.44&1.52&0.94\\
&$\varepsilon_{e\tau}$&0.23&1.54&1.54&2.9\\\hline
\end{tabular}
\label{table:chi2}
\end{table}

\begin{figure*}[t!]
\vspace*{-0.5cm}
\hspace*{-0.1cm}
\includegraphics[height=5.87cm,width=5.87cm]{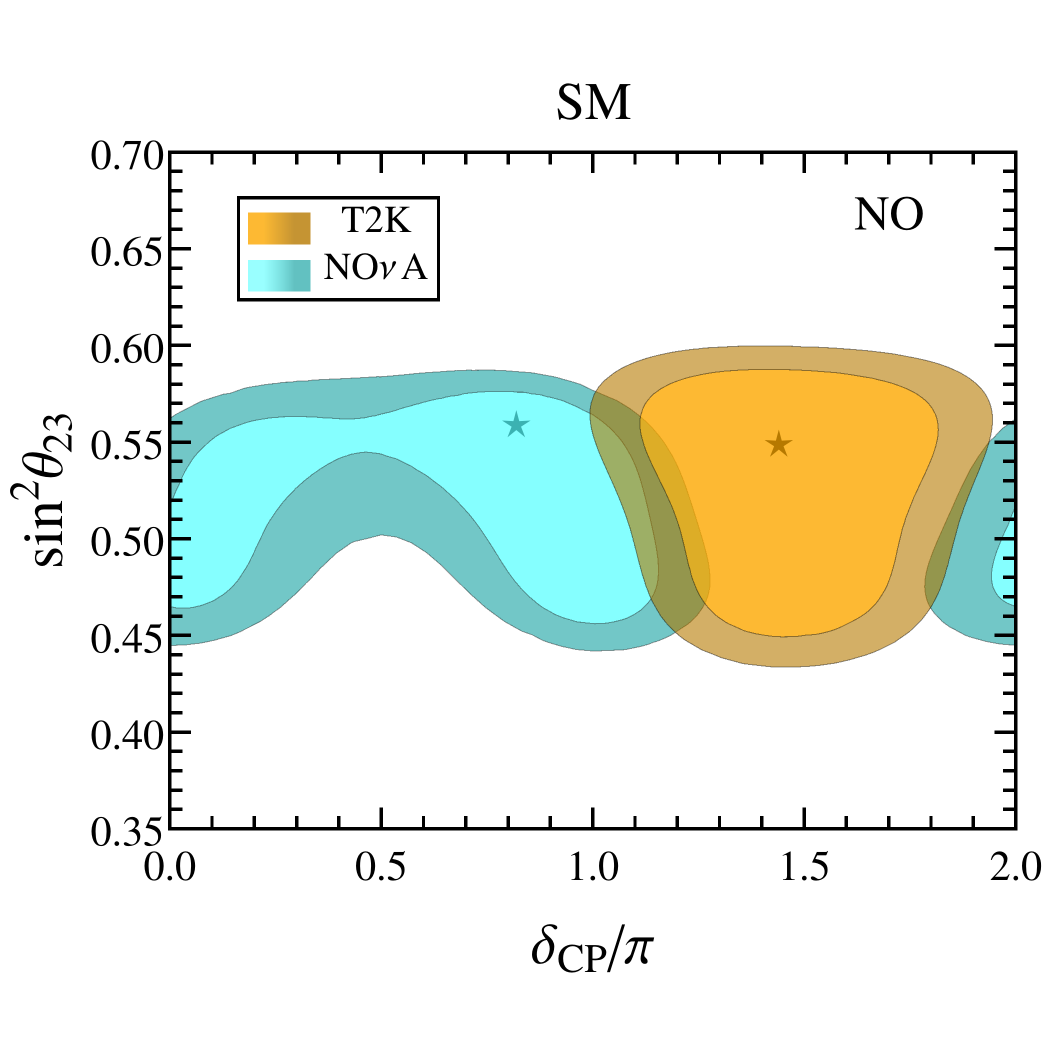}
\includegraphics[height=5.87cm,width=5.87cm]{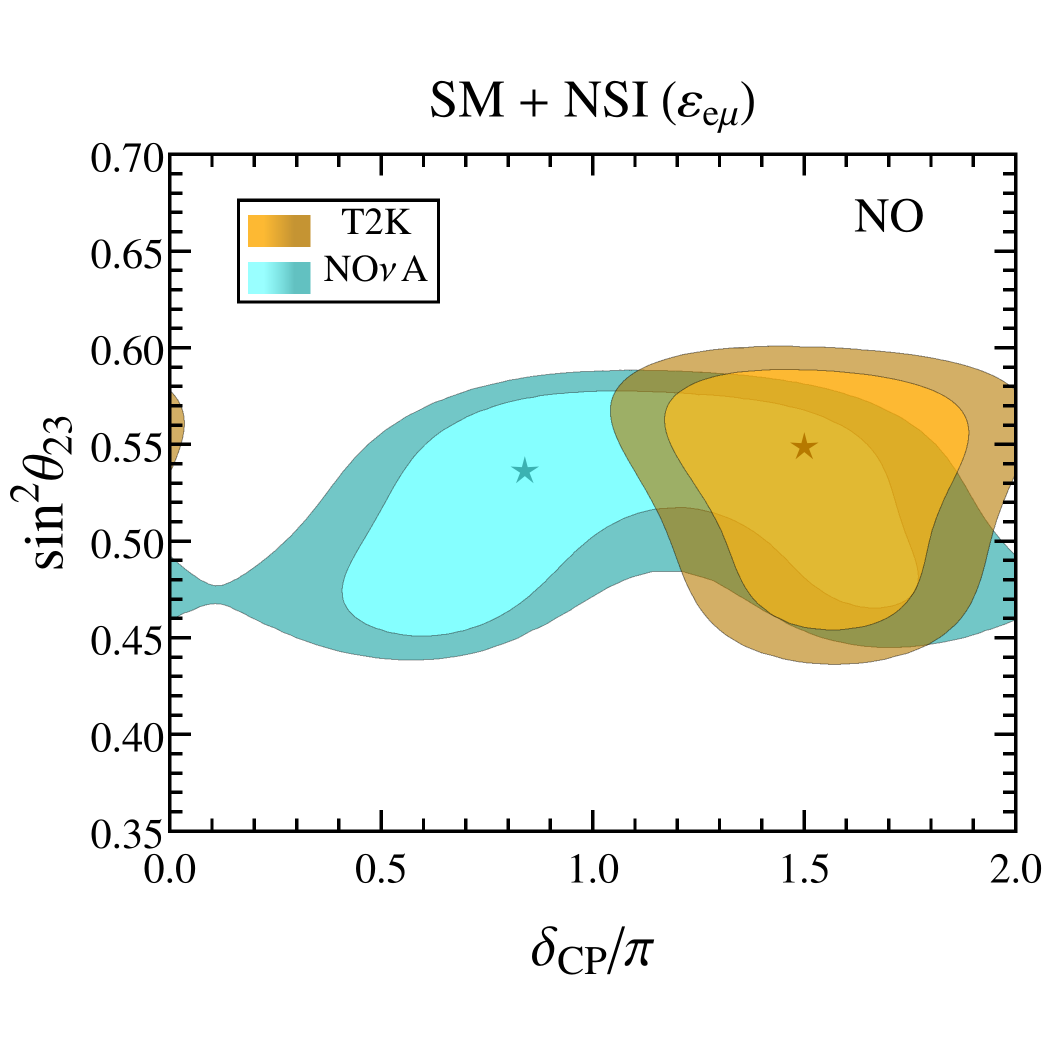}
\includegraphics[height=5.87cm,width=5.87cm]{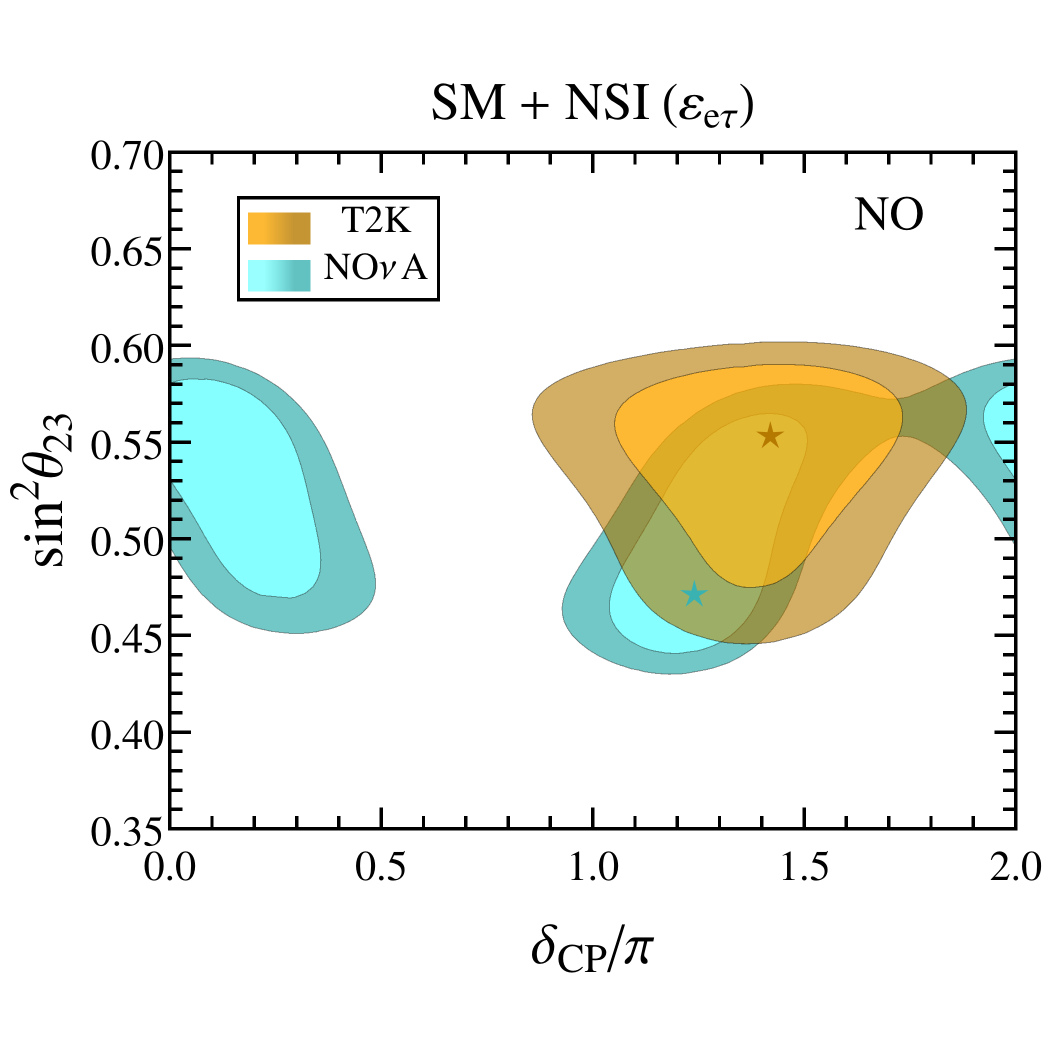}
\vspace*{-0.3cm}
\caption{Allowed regions by T2K and NOvA for NO in the SM case
(left panel) and with NSI of the $e-\mu$ type (central panel) and of the $e-\tau$ type
 (right panel). In the central panel we have taken 
the NSI parameters at their best fit values of the {\em{combination}} 
T2K + NOvA. These correspond to ($|\varepsilon_{e\mu}| = 0.125, \phi_{e\mu} = 1.35\pi$)
for the central panel and ($|\varepsilon_{e\tau}| = 0.22, \phi_{e\tau} = 1.70\pi$) for the right panel.
The contours correspond to the 68$\%$ and 90$\%$ C.L. for 2 d.o.f.}
\label{fig_tension_NO}
\end{figure*} 

\begin{figure*}[t!]
\vspace*{-0.0cm}
\hspace*{-0.1cm}
\includegraphics[height=5.87cm,width=5.87cm]{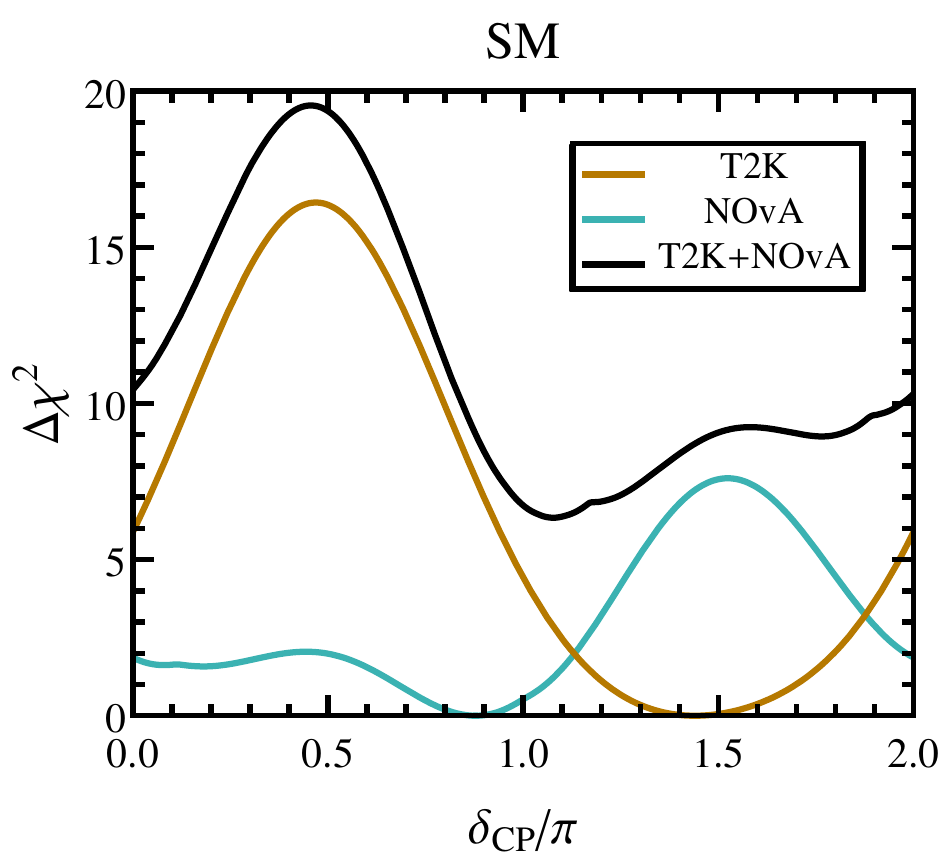}\includegraphics[height=5.87cm,width=5.87cm]{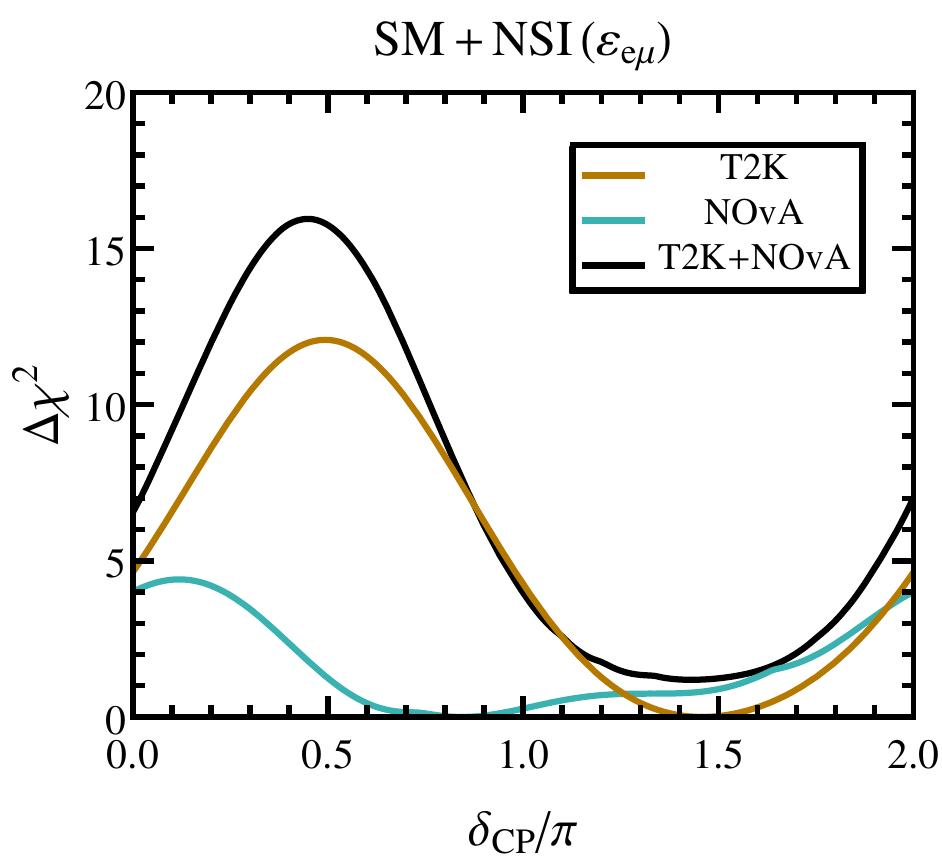}
\includegraphics[height=5.87cm,width=5.87cm]{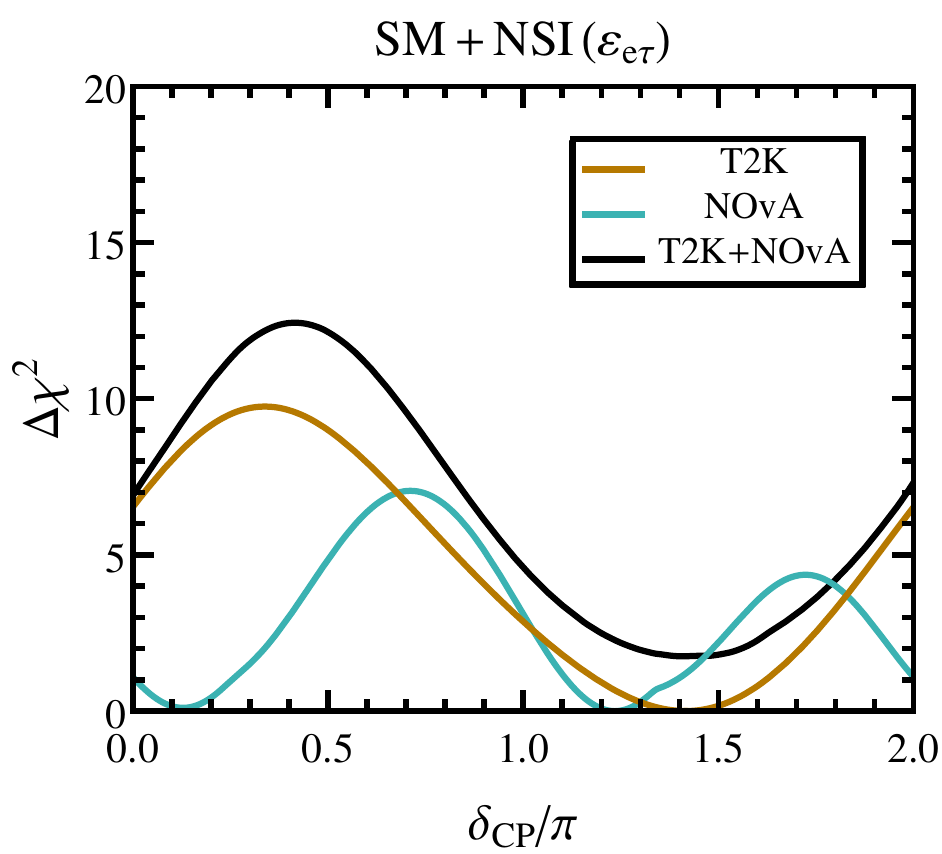}
\vspace*{-0.3cm}
\caption{Plot of the functions $\Delta\chi^2_{\rm{T2K}}$, $\Delta\chi^2_{\rm{NOvA}}$ and  $\bar\chi^2$ 
 as a function of $\delta_{\mathrm {CP}}$ for normal ordering. The left panel corresponds to the SM case (same as Fig.~1),
 the other two panels represent the two NSI cases of $e-\mu$ type (central panel) and  $e-\tau$ type
 (right panel). In the central panel we have taken 
the NSI parameters at their best fit values of the {\em{combination}} 
T2K + NOvA as in Fig.~\ref{fig_tension_NO}.}
\label{fig_tension_NO_1D}
\end{figure*} 

\begin{figure*}[t!]
\vspace*{-0.0cm}
\hspace*{-0.2cm}
\includegraphics[height=6.2cm,width=6.2cm]{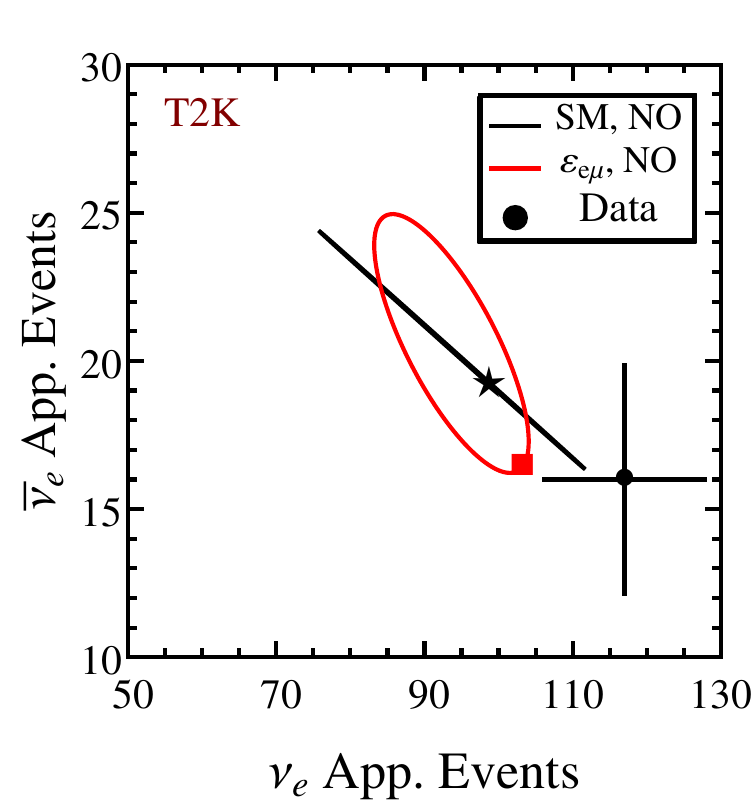}
\includegraphics[height=6.2cm,width=6.2cm]{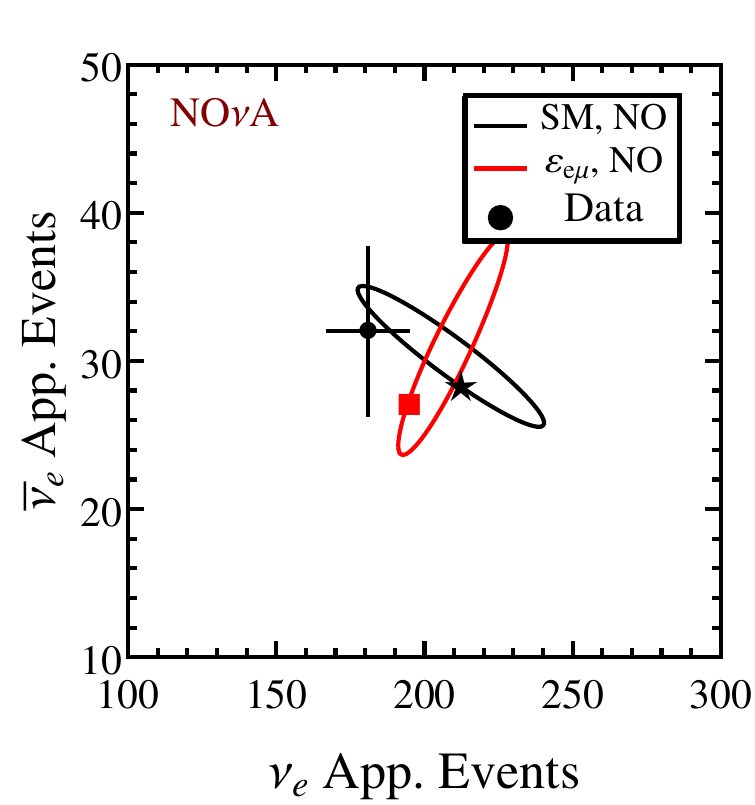}
\includegraphics[height=6.2cm,width=6.2cm]{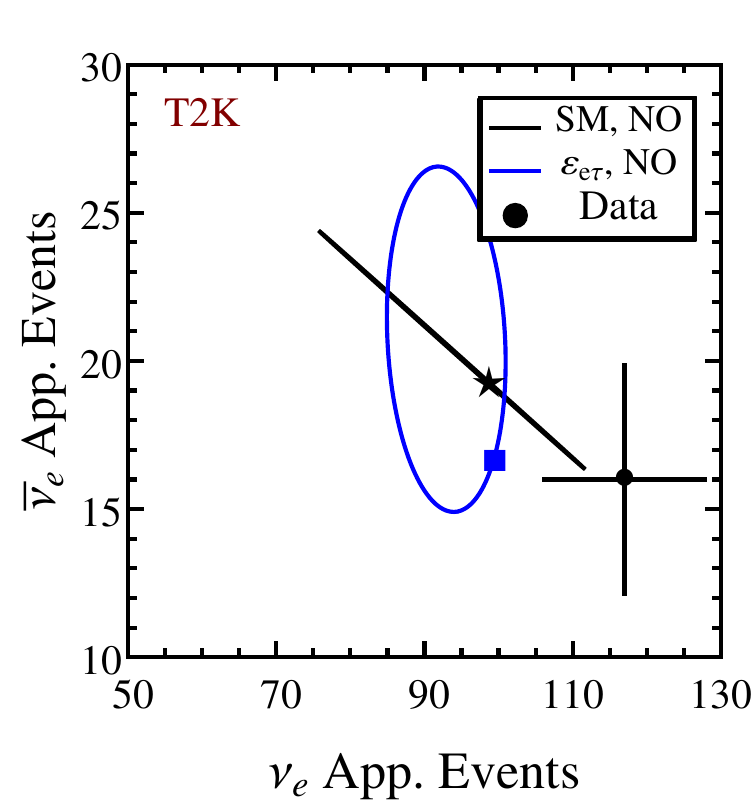}
\includegraphics[height=6.2cm,width=6.2cm]{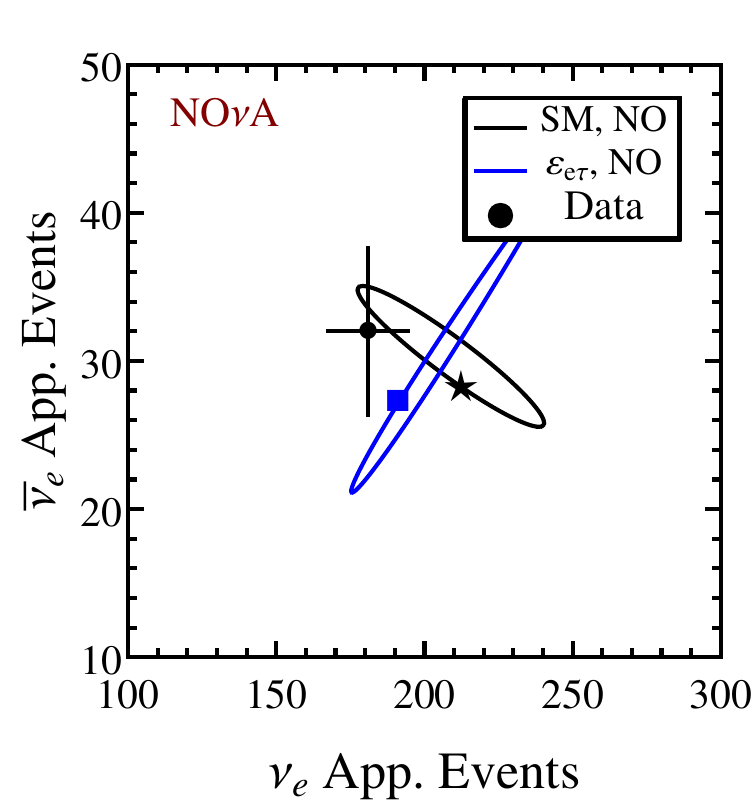}
\caption{Bievents plots for the T2K (left panels) and NOvA setup (right panels) for the NO case.
The upper (lower) panels represent the case of $\varepsilon_{e\mu} (\varepsilon_{e\tau}$).
Along all the ellipses the running parameter is the standard CP phase $\delta_{\rm CP}$ in the
range $[0, 2\pi]$. The black ellipses correspond to the SM case with best fit represented by stars.
The colored ellipses correspond to the SM+NSI case with best fit indicated by squares. 
The ellipses and the best fit points located on them are determined by fitting the {\em combination} 
of T2K and NOvA. The points with the error bars display the experimental 
data with their statistical uncertainties.}
\label{fig:bievents-plot_NO}
\end{figure*} 

In Fig.~\ref{fig_tension_NO_1D}  we shows the function $\bar \chi^2(\delta_{\mathrm {CP}})$
together with the two functions  $\Delta\chi^2_{r}(\delta_{\mathrm {CP}}) = \chi^2_{r}(\delta_{\mathrm {CP}}) - \chi^2_{r,min }$,
where $r$ is an index designating the experiment in question (T2K or NOvA). The left panel coincides
with Fig.~\ref{fig:chi2-plot_NO}, which is reported here again for the sake of clearness, in order to facilitate
the visual comparison with the cases corresponding to NSI presented in the central  panel ($\varepsilon_{e\mu}$ case)
and the right one ($\varepsilon_{e\tau}$ case). From these last two panels it is clear how in the presence
of NSI the level of tension is substantially reduced and it is basically negligible.%
\footnote{In this plot we have decided to show the $\Delta\chi^2$ curves for T2K and NOvA for the best fit
of their {\em{combination}} in order to facilitate the comparison of Fig.~\ref{fig_tension_NO_1D} with Fig.~\ref{fig_tension_NO}.
Indeed, with this choice, the $\Delta\chi^2$  curves are exactly the 1D projections of the 2D regions of Fig.~\ref{fig_tension_NO}.
Note, however that for correctly estimating the GoF one should consider (not shown) the curves corresponding to T2K and NOvA  
{\em{marginalizing}} over the NSI parameters  ($\varepsilon_{e\mu}$ and  $\phi_{e\mu}$ in the central panel and
$\varepsilon_{e\tau}$ and  $\phi_{e\tau}$ in the right panel).  Clearly, in this case, there are 4 d.o.f
(two oscillation parameters and two NSI parameters). Following this procedure we find $\bar\chi^2_{min} = 4.1$ 
for the $e-\mu$ case and $\bar\chi^2_{min} = 4.5$ for $e-\tau$ case.
Considering 4 d.o.f. these values correspond to a ${\mathrm{GoF}} = 3.9 \times 10^{-1}$
and ${\mathrm{GoF}} = 3.4 \times 10^{-1}$ respectively, which are both very high.}

As a further tool to interpret the results of the analysis, in Fig.~\ref{fig:bievents-plot_NO}, 
we show the bievents plots, in which on the $x$-axis ($y$-axis) is reported the number 
of detected electron neutrino (antineutrino) events. Such graph is of guide to understanding 
the source of the tension between NOvA and T2K
in the SM scenario and its alleviation in the presence of NSI. The plot
is particularly insightful because in NOvA and T2K, almost all the information 
given in the appearance channel data is condensed in the number of events collected. 
In fact, due to the low statistics, the information contained in the shape of the energy spectrum is still limited.
The ellipses displayed in the figure are plotted using the best fit parameters of the {\em combination} of T2K and NOvA. 
The relevant parameters  are
$\theta_{23}$, $\theta_{13}$ and $\Delta m^2_{31}$ for the SM case. For the SM+NSI framework,
one has also $|\varepsilon_{\alpha\beta}|$ and $\phi_{\alpha\beta}$. 
Both in the SM and SM+NSI cases, the running parameter along the ellipses is the CP 
phase $\delta_{\rm CP}$ in the range $[0, 2\pi]$. The black ellipses correspond to the SM scenario 
with the stars representing the best fit point $\delta_{\rm CP}^{\rm SM} = 1.08\pi$. Such a value of
 $\delta_{\rm CP}$ is a compromise among T2K (which push towards  $\delta_{\rm CP} = 1.5\pi$) and NOvA (which 
tends to prefer values close to $0.9\pi$).  The colored ellipses correspond the SM+NSI scenario
 (the squares designate the best fit value $\delta_{\rm CP}^{\rm NSI} \simeq 1.4 \pi$, which is approximately
 the same in both the $e-\mu$ and $e-\tau$ options). 
 The upper (lower) panels represent to the $e-\mu$ ($e-\tau$) scenario. From the plots it is well visible how in the presence
 of the NSI, the best fit point of the model gets closer to the experimental data, thus reducing the tension found in the SM case.
 For completeness, in the  Supplemental Material
we provide the additional figures~S1-S5.

{\bf {\em Conclusions.}} In this paper we have reassessed the issue of the tension 
between the measurements performed in the appearance channel by 
T2K and NOvA. The discrepancy first emerged at the Neutrino 2020 conference
persists in the latest data released at Neutrino 2024 conference. We find that the disagreement
can be resolved by non-standard interactions (NSI) of the flavor changing type involving the
$e-\mu$ and $e-\tau$ flavors. Further experimental information is needed in order to settle the issue. The next data expected 
to come from T2K and NOvA will be crucial in this respect. Also, complementary information
may be extracted from ANTARES, IceCube  and KM3NeT/ORCA experiments which are sensitive to the 
relevant NSI couplings. Most probably, however, if the current indication in favor of  NSI will persist,
only the future new-generation experiments DUNE and Hyper-Kamiokande will be able
to definitely (dis)confirm it.%
\footnote{Note that these experiments may provide information on NSI also by observing astrophysical signals
like those expected to come from supernova neutrinos (see for example~\cite{Jana:2024lfm}).}

Our results indicate the presence of effective NSI couplings of the order of ten percent. 
Considering Eq.~(\ref{epsilon_eff}), these may translate into couplings of a few per 
cent for the fundamental particles ($u$ and $d$ quarks and electrons). 
Still, these are quite large couplings from a theoretical standpoint.  In fact, if NSI are 
induced by mediators heavier than the electroweak 
scale, one naturally finds that the charged leptons are sensitive to
new physics, on which there are strong limits. One possible way to avoid this issue is to augment the complexity
of the model by considering dimension-8 operators~\cite{Gavela:2008ra},  invoking radiative
neutrino mass models (see for example the recent studies~\cite{Babu:2019mfe,Forero:2016ghr,Dey:2018yht}), or
calling in to play vector leptoquarks~\cite{Majhi:2022wyp}. A radically different and fascinating option is to consider 
NSI induced by light or ultra-light mediators, which are gaining increasing attention 
(see for example~\cite{Farzan:2015doa,Farzan:2015hkd}). In this case, the NSI
effects, due to the low momentum transfer, will be hardly visible in
processes other than neutrino oscillations. As a matter of fact, the coherent forward scattering of 
neutrinos with ambient particles, operative at zero momentum transfer and observable through
the modification of the oscillation probabilities, would provide the only way to probe new physics beyond SM
in the neutrino sector.

We hope that our study may trigger investigations
 both at the experimental level, deepening the understanding
of systematic uncertainties, and on the theoretical side, seeking new models able to predict the preferred 
NSI couplings.


\begin{acknowledgments}

\noindent  We thank Enrique Fern\'andez-Mart\'inez and Thomas Schwetz for useful discussions. The work of S.S.C. is funded by the Deutsche Forschungsgemeinschaft (DFG, German Research Foundation) – project number 510963981. A.P. acknowledges partial support by the research grant number 2022E2J4RK ``PANTHEON: Perspectives in Astroparticle and Neutrino THEory with Old and New messengers" under the program PRIN 2022 funded by the Italian Ministero dell’Universit\`a e della Ricerca (MUR) and by the European Union – Next Generation EU as well as partial support by the research project {\em TAsP} funded by the Instituto Nazionale di Fisica Nucleare (INFN).

\end{acknowledgments}

\bibliographystyle{utphys}
\bibliography{NSI-References_v3}

\clearpage
\newpage
\maketitle
\onecolumngrid
\begin{center}
\textbf{\large Supplemental Material} \\ 
\vspace{0.05in}

\end{center}

\setcounter{equation}{0}
\setcounter{figure}{0}
\setcounter{table}{0}
\setcounter{section}{1}
\renewcommand{\theequation}{S\arabic{equation}}
\renewcommand{\thefigure}{S\arabic{figure}}
\renewcommand{\thetable}{S\arabic{table}}
\newcommand\ptwiddle[1]{\mathord{\mathop{#1}\limits^{\scriptscriptstyle(\sim)}}}

\begin{figure*}[h!]
\vspace*{-0.0cm}
\hspace*{-0.2cm}
\includegraphics[height=4.9cm,width=4.9cm]{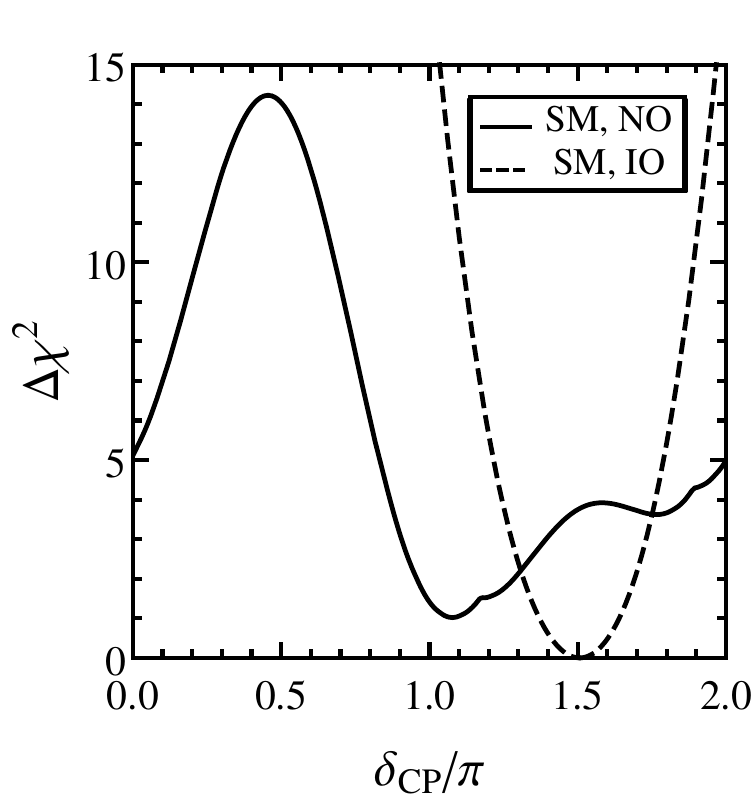}
\includegraphics[height=4.9cm,width=4.9cm]{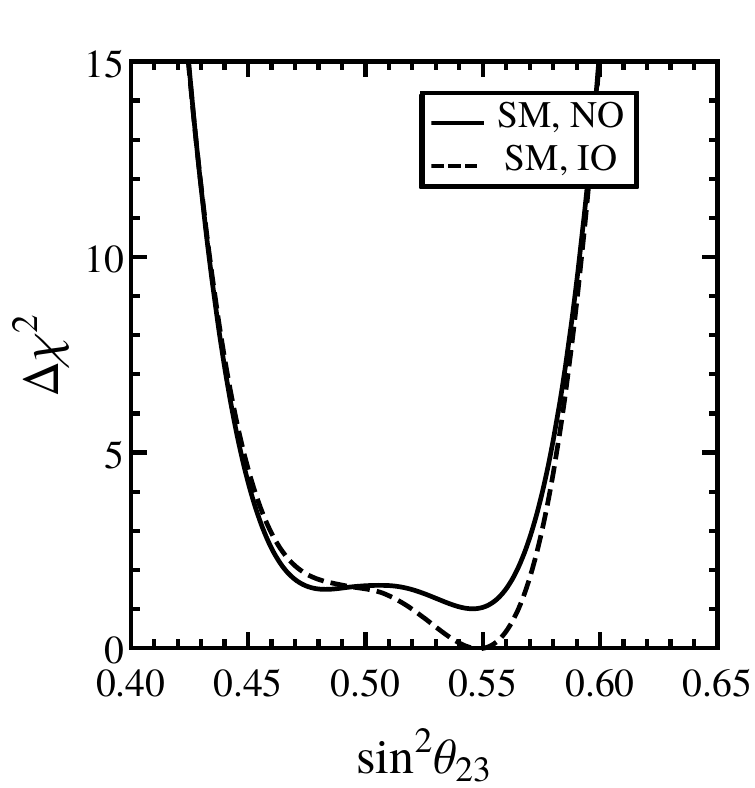}
\includegraphics[height=4.9cm,width=4.9cm]{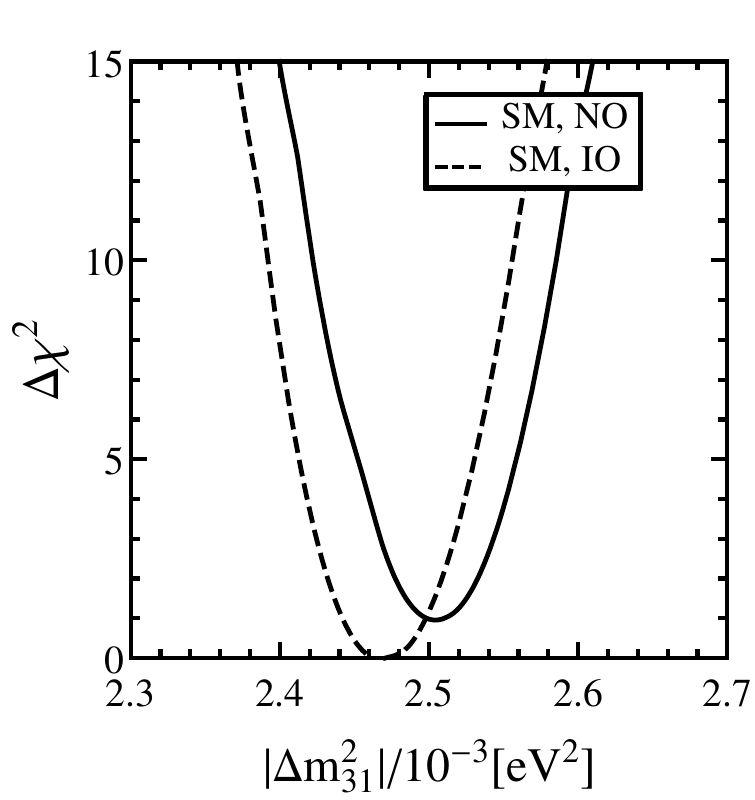}\\
\includegraphics[height=4.9cm,width=4.9cm]{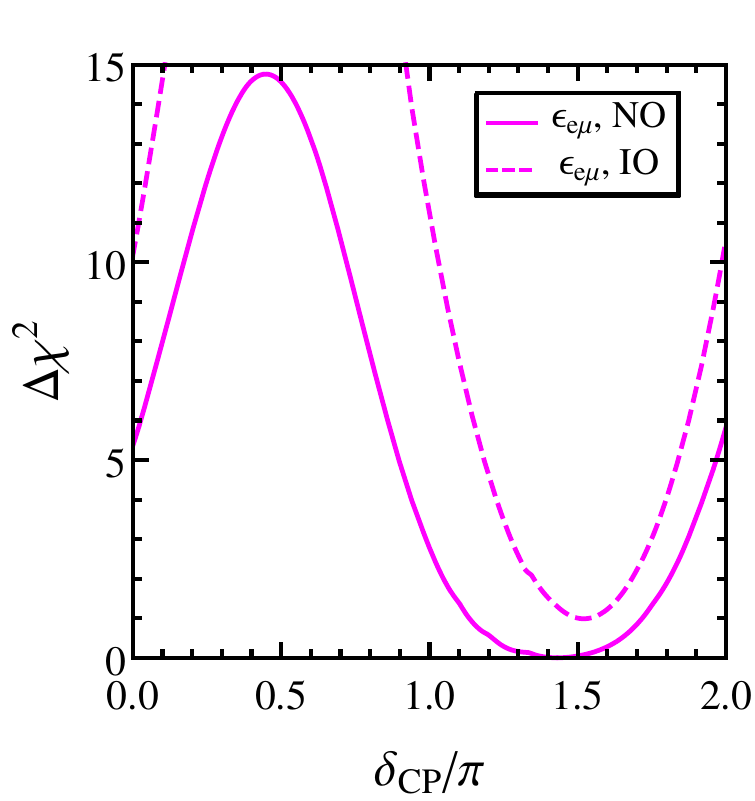}
\includegraphics[height=4.9cm,width=4.9cm]{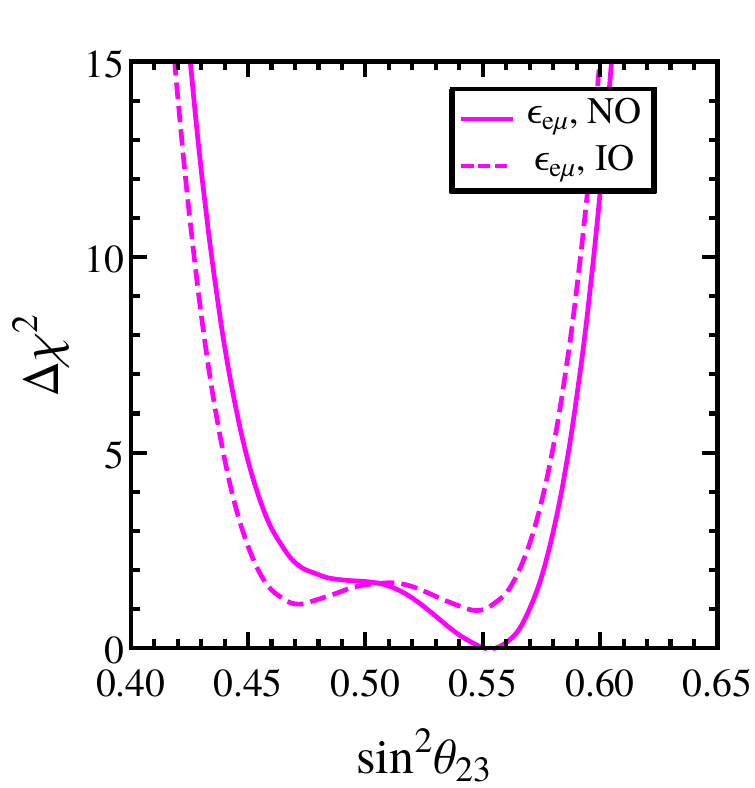}
\includegraphics[height=4.9cm,width=4.9cm]{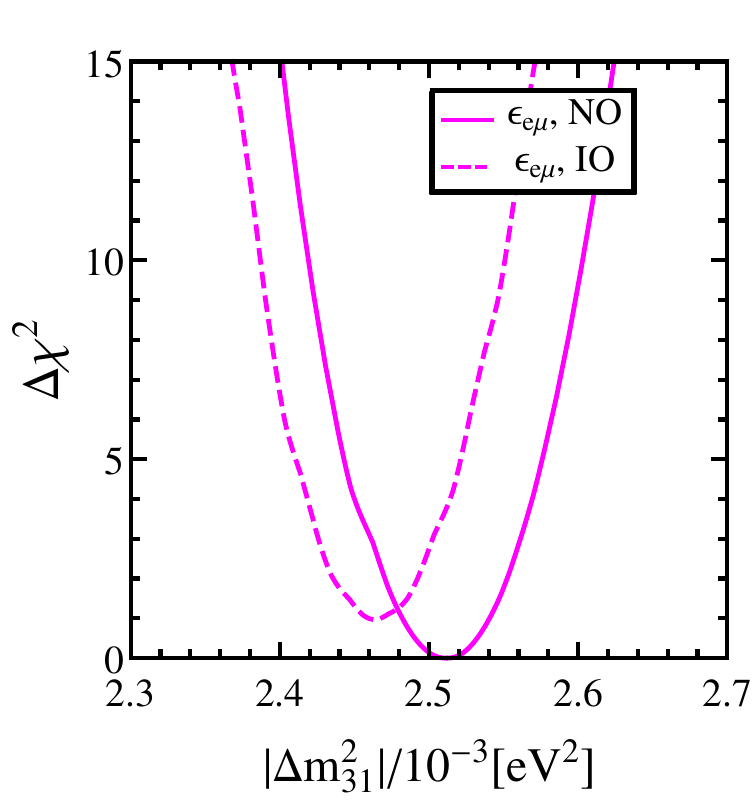}\\
\includegraphics[height=4.9cm,width=4.9cm]{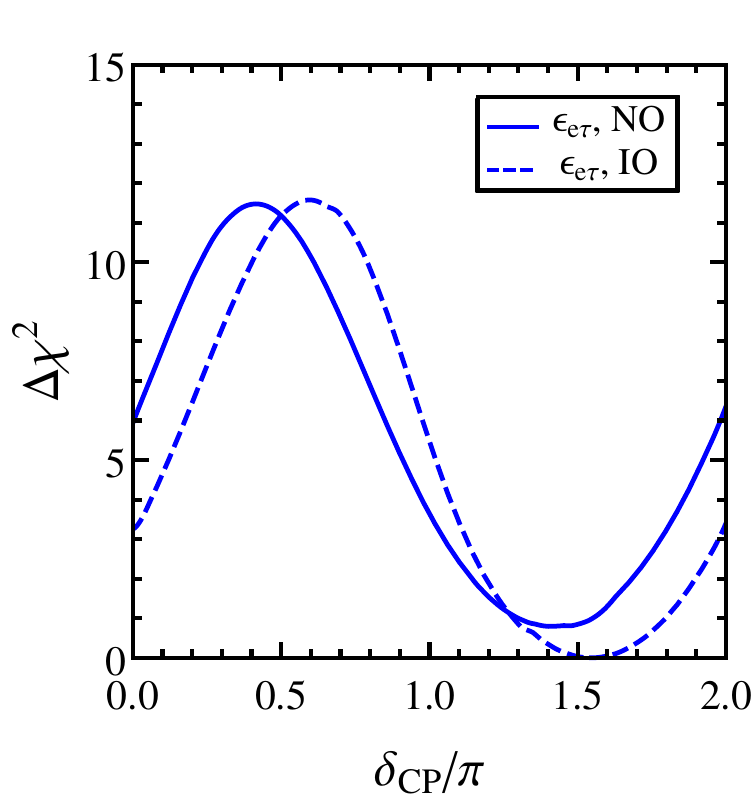}
\includegraphics[height=4.9cm,width=4.9cm]{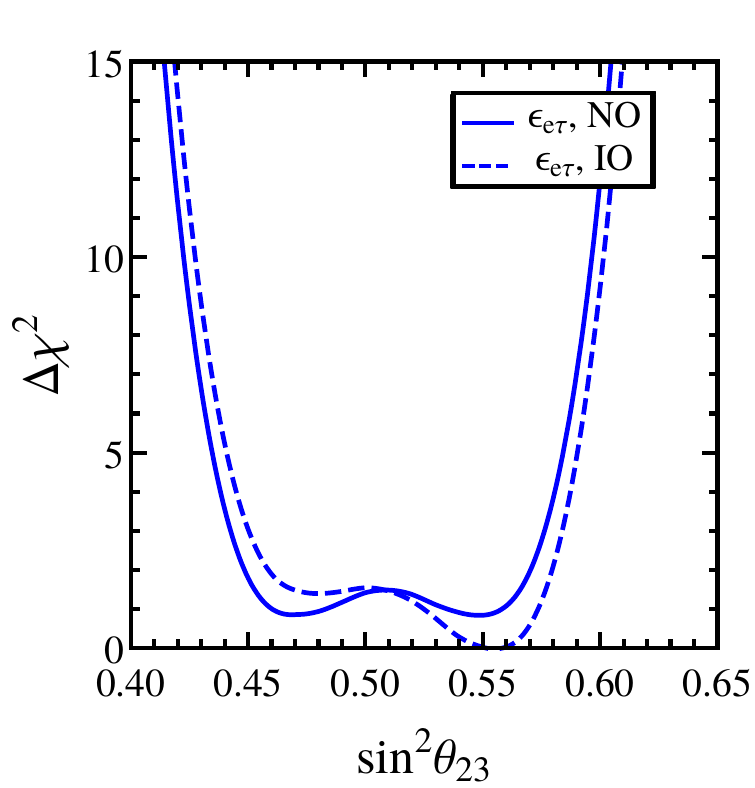}
\includegraphics[height=4.9cm,width=4.9cm]{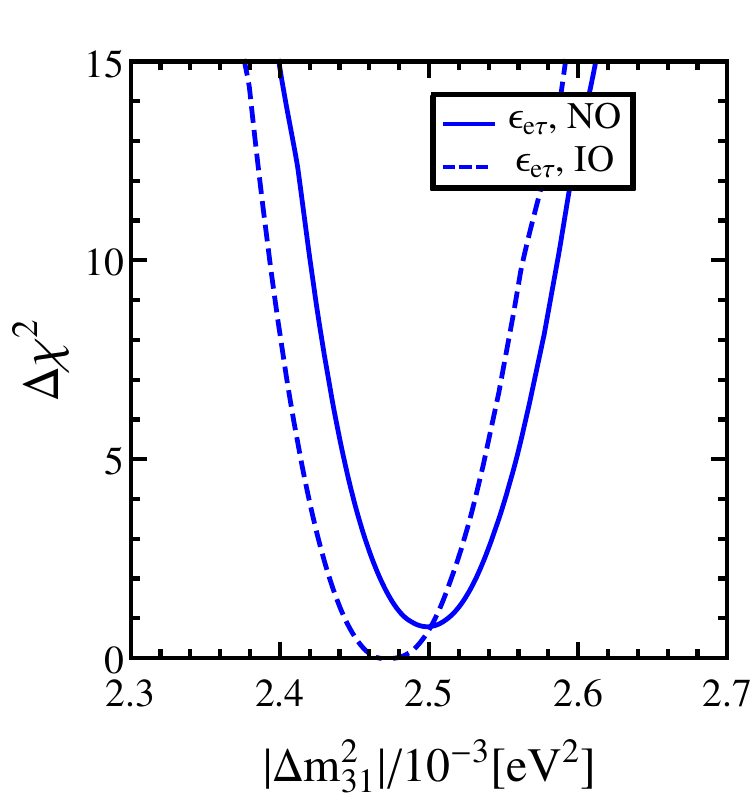}
\vspace*{-0.0cm}
\caption{One-dimensional projections of $\Delta \chi^2$ on the standard parameters determined by the combination of T2K and NO$\nu$A for NO (continuous curves) and IO (dashed curves).}
\label{fig:regions_1D}
\end{figure*}

\begin{figure}[b!]
\vspace*{-0.0cm}
\hspace*{-0.0cm}
\includegraphics[height=10.5cm,width=10.5cm]{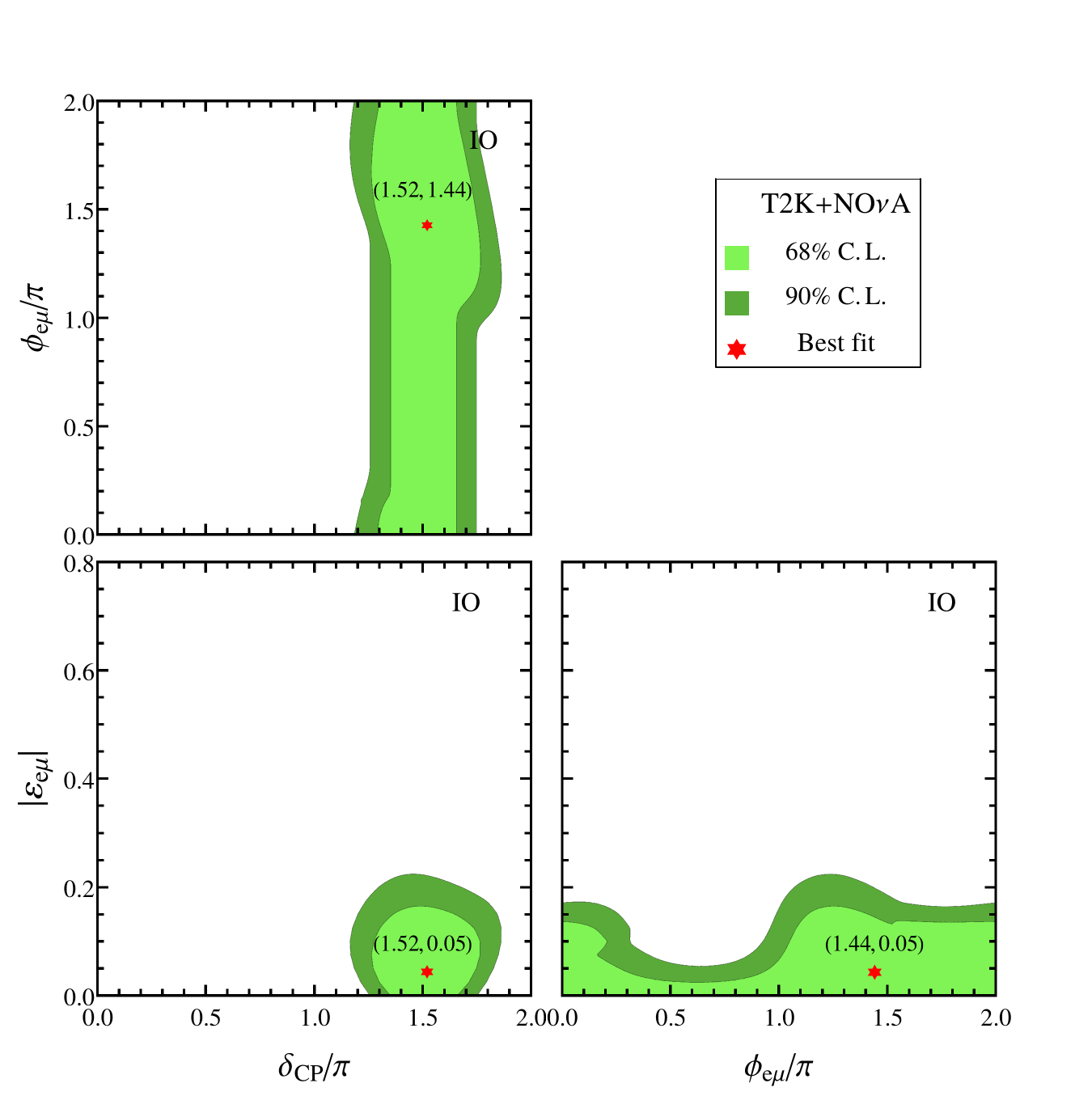}
\caption{Allowed regions determined by the combination of T2K and NO$\nu$A in IO for NSI of the $e-\mu$ type.}
\label{fig:regions_emu_IO}
\end{figure}

\begin{figure}[b!]
\vspace*{-0.0cm}
\hspace*{-0.0cm}
\includegraphics[height=10.5cm,width=10.5cm]{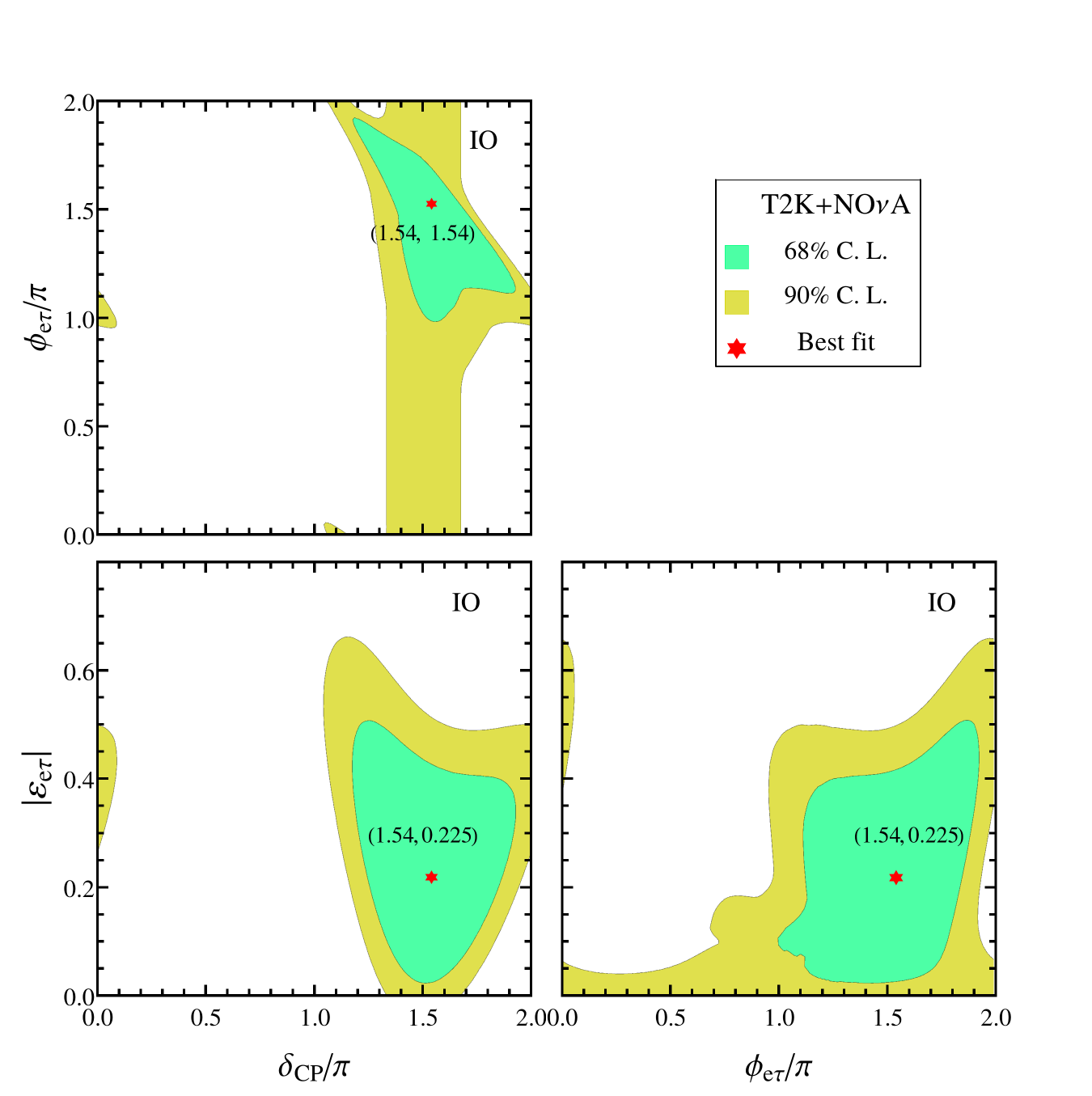}
\caption{Allowed regions determined by the combination of T2K and NO$\nu$A in IO for NSI of the $e-\tau$ type.}
\label{fig:regions_etau_IO}
\end{figure}

\begin{figure*}[h!]
\vspace*{-0.0cm}
\hspace*{-0.1cm}
\includegraphics[height=5.87cm,width=5.87cm]{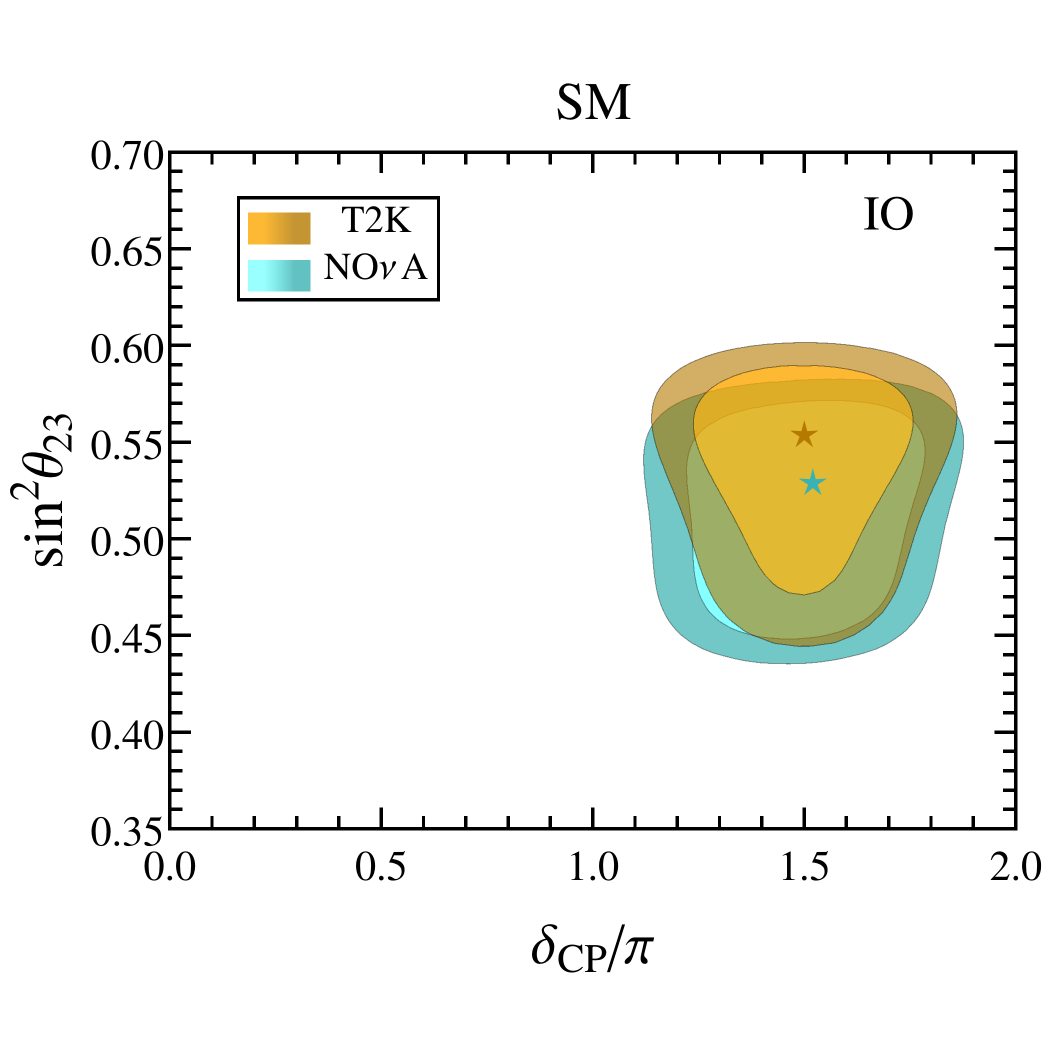}
\includegraphics[height=5.87cm,width=5.87cm]{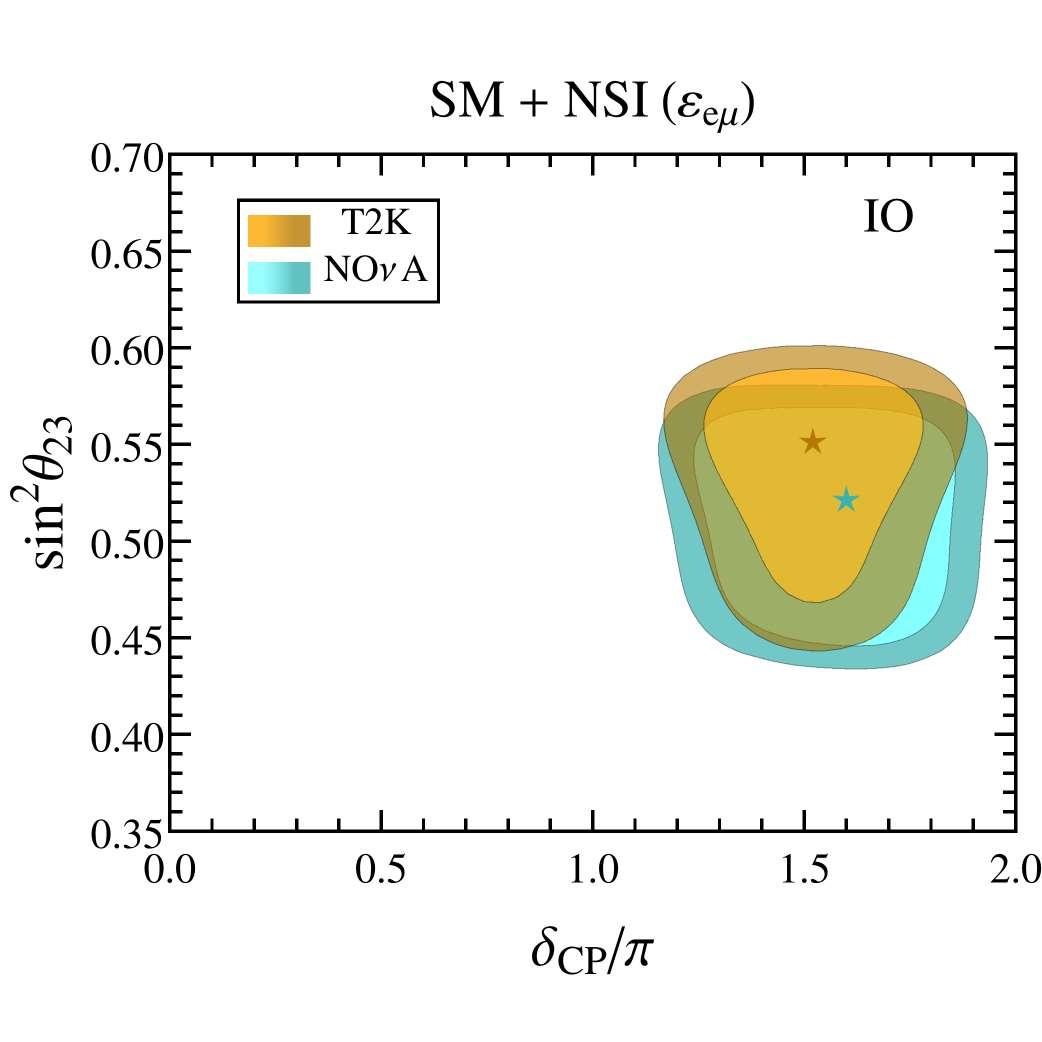}
\includegraphics[height=5.87cm,width=5.87cm]{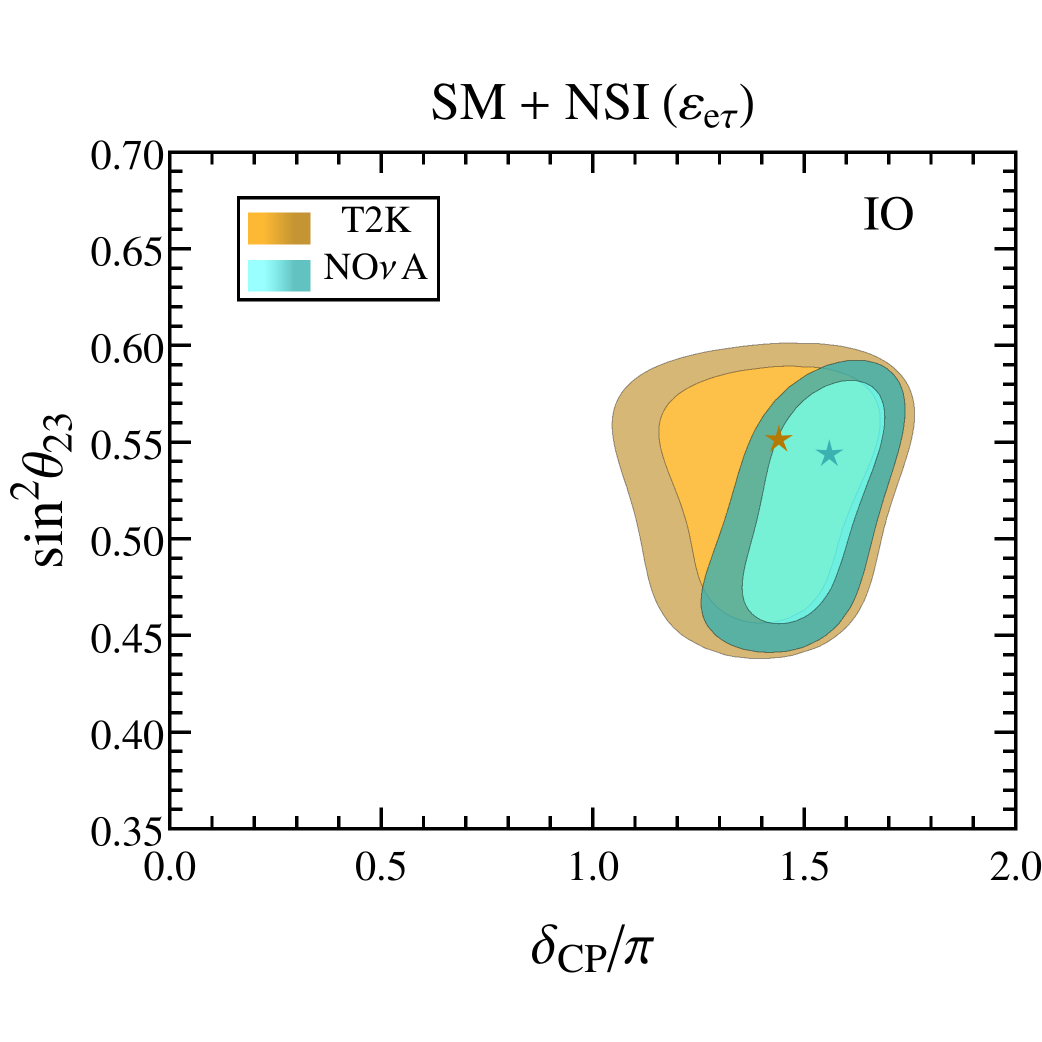}
\vspace*{-0.5cm}
\caption{Allowed regions determined separately by T2K and NO$\nu$A for IO in the SM case
(left panel) and with NSI in the $e-\mu$ sector (middle panel) and in the $e-\tau$ sector
 (right panel). In the middle panel we have taken 
the NSI parameters at their best fit values of T2K + NO$\nu$A ($|\varepsilon_{e\mu}| = 0.05, \phi_{e\mu} = 1.44\pi$).
Similarly, in the right panel we have taken $|\varepsilon_{e\tau}| = 0.23, \phi_{e\tau} = 1.54\pi$.
The contours are drawn at the 68$\%$ and 90$\%$ C.L. for 2 d.o.f.}
\label{fig_tension_IO}
\end{figure*}

\begin{figure}[b!]
\vspace*{-0.0cm}
\hspace*{-0.2cm}
\includegraphics[height=6.2cm,width=6.2cm]{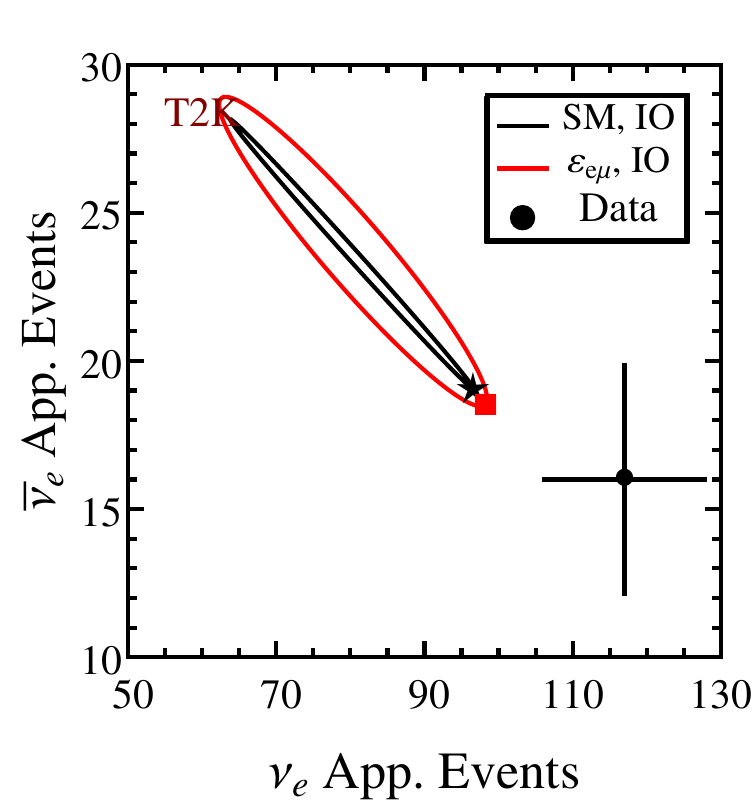}
\includegraphics[height=6.2cm,width=6.2cm]{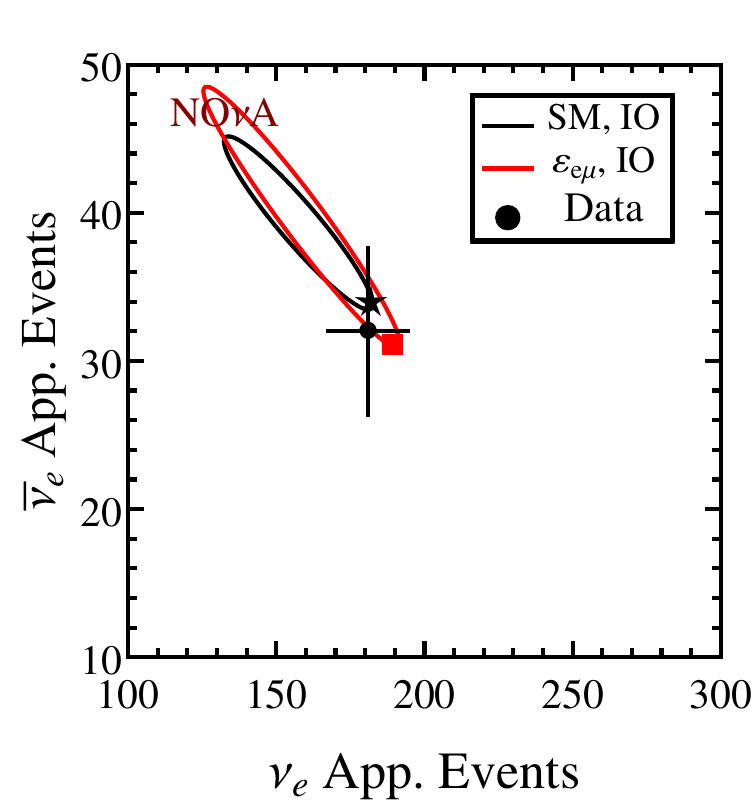}
\includegraphics[height=6.2cm,width=6.2cm]{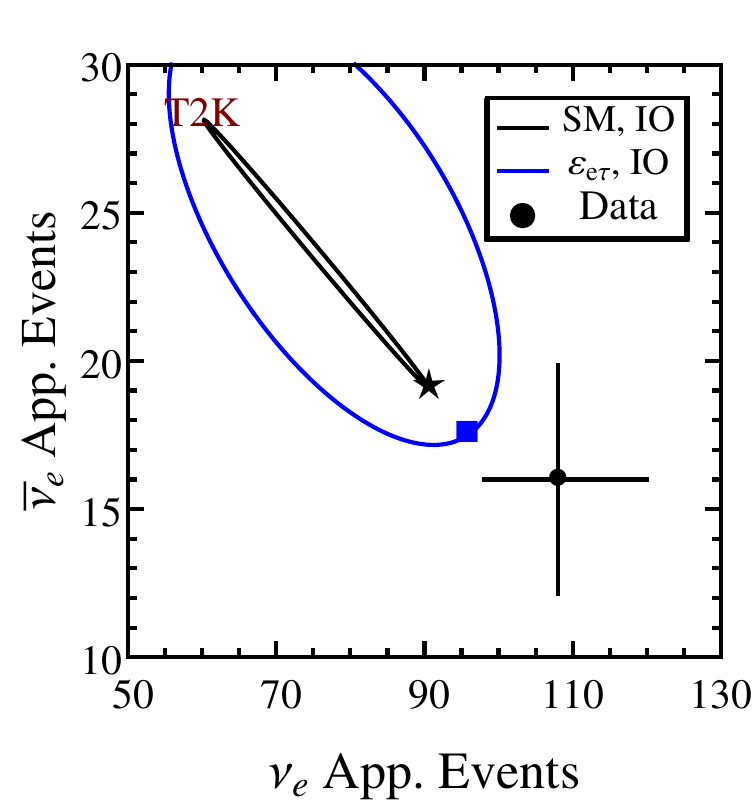}
\includegraphics[height=6.2cm,width=6.2cm]{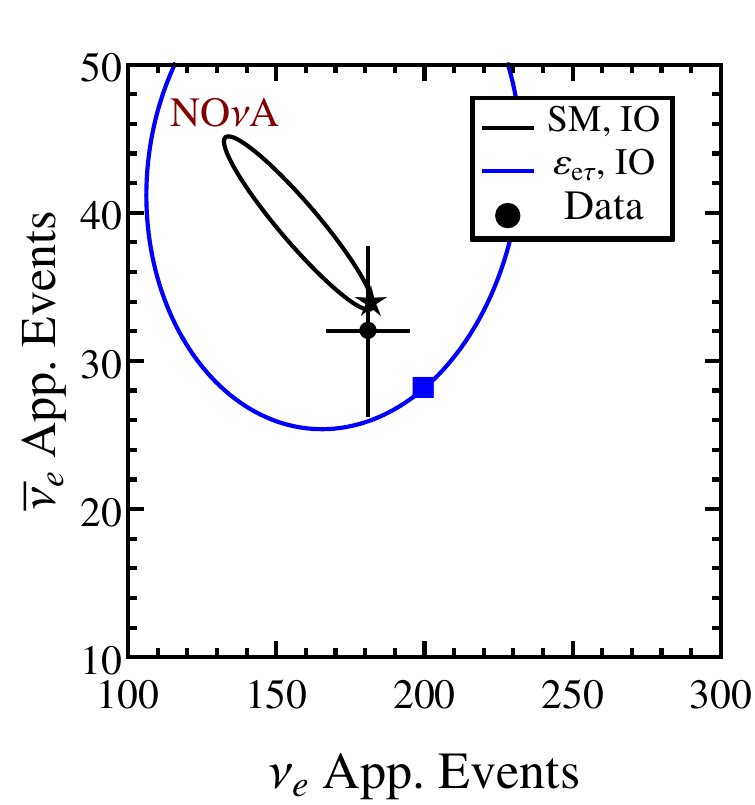}
\caption{Bievents plots for the T2K (left panels) and NO$\nu$A setup (right panels) for the IO case.}
\label{fig:bievents-plot_IO}
\end{figure}

\noindent In the Supplemental Material, we present  five additional figures. 
In Fig.~\ref{fig:regions_1D} we show the one-dimensional projections of $\Delta \chi^2$ on the standard 
oscillation parameters $\delta_{\mathrm {CP}}$, $\theta_{23}$ and $|\Delta m^2_{31}|$
from the combination of NO$\nu$A and T2K, with and without NSI. 
Fig~\ref{fig:regions_emu_IO} and Fig.~\ref{fig:regions_etau_IO} present the allowed
regions obtained from the analysis in the case of IO for both $e-\mu$ and $e-\tau$ couplings.
Fig.~\ref{fig_tension_IO} shows the allowed regions for the IO case, 
in the plane of the two parameters $\delta_{\mathrm {CP}}$ and $\theta_{23}$.
Finally, in Fig.~\ref{fig:bievents-plot_IO} we present the bievents plots for the IO case.

\end{document}